\documentclass[journal]{IEEEtran}

\IEEEoverridecommandlockouts

\usepackage[utf8]{inputenc}
\usepackage{cite}
\usepackage{amsmath,amssymb,amsfonts,amsthm}
\usepackage{algorithmic}
\usepackage{graphicx}
\usepackage{textcomp}
\usepackage{xcolor}
\usepackage{booktabs}
\usepackage{multirow}
\usepackage{array}
\usepackage{url}
\usepackage{hyperref}
\usepackage{balance}
\usepackage{threeparttable}
\usepackage[framemethod=default]{mdframed}
\usepackage{enumitem}
\usepackage{placeins}

\definecolor{promptbg}{gray}{0.95}
\definecolor{promptrule}{gray}{0.65}
\newmdenv[
  backgroundcolor=promptbg,
  linecolor=promptrule,
  linewidth=0.4pt,
  innerleftmargin=8pt,
  innerrightmargin=8pt,
  innertopmargin=6pt,
  innerbottommargin=6pt,
  skipabove=4pt,
  skipbelow=6pt
]{promptbox}

\graphicspath{{figures/}}

\begin{document}

\title{SemABR: Measuring Video Semantic Fidelity with Multimodal LLMs for Adaptive Bitrate Streaming}

\author{
  Shiqi~Xu,
  Soung Chang~Liew,~\IEEEmembership{Life Fellow,~IEEE},
  Yuyang~Du,~\IEEEmembership{Member,~IEEE}
  
  \thanks{The authors are with the Department of Information Engineering, The Chinese University of Hong Kong, Hong Kong SAR, China (e-mail: \{xs024, soung, yuydu\}@ie.cuhk.edu.hk). Corresponding author: Soung Chang Liew.}
  \thanks{The work was supported in part by the Hong Kong Innovation and Technology Fund (Project Number: ITS/362/24). The experimental work in this paper was conducted in the JC STEM Lab of Advanced Wireless Networks for Mission-Critical Automation and Intelligence funded by The Hong Kong Jockey Club Charities Trust.}
}

\maketitle

\begin{abstract}
Conventional video metrics such as PSNR, SSIM, and VMAF measure visual distortion or perceptual quality, but they do not directly capture semantic preservation: whether compression retains a video's objects, actions, and temporal narrative. This gap matters for adaptive bitrate (ABR) streaming because user experience depends not only on visual quality and playback smoothness but also on whether such semantic information is preserved. Existing Quality-of-Experience (QoE)-driven bitrate-selection and resource-allocation methods primarily aim to minimize rebuffering and bitrate switching while maximizing perceptual video quality, without explicitly considering semantic preservation. To address this gap, we introduce video semantic fidelity (SF), a metric that quantifies how well a compressed video preserves the semantic content of its source. An offline multimodal large language model (MLLM) generates structured descriptions of the reference and compressed versions of the video, and a separate text-only large language model (LLM) evaluates their semantic correspondence. The resulting content-dependent SF--bitrate profiles are cached and queried by the online bitrate selector without invoking MLLMs at runtime. Evaluations on three subjective QoE benchmarks show a consistent positive association between SF and mean opinion scores (MOS). A separate human semantic-rating study directly evaluates semantic preservation and shows that SF correlates more strongly with human judgments than conventional video metrics. We then embed these profiles into a 5G MEC-assisted video-on-demand (VoD) resource-allocation framework at the base station. When wireless resources cannot support high bitrate levels for all users, the framework uses the SF--bitrate profiles to select bitrate levels jointly across users and reduce the semantic loss caused by the required bitrate reductions. NS-3 simulations with the 5G NR module show that the proposed framework achieves higher average and worst-user SF than the evaluated baselines, with a widening advantage as the wireless resources available to each user decrease.
\end{abstract}


\section{Introduction}
\label{sec:intro}

\begin{figure}[!t]
  \centering
  \includegraphics[width=\columnwidth]{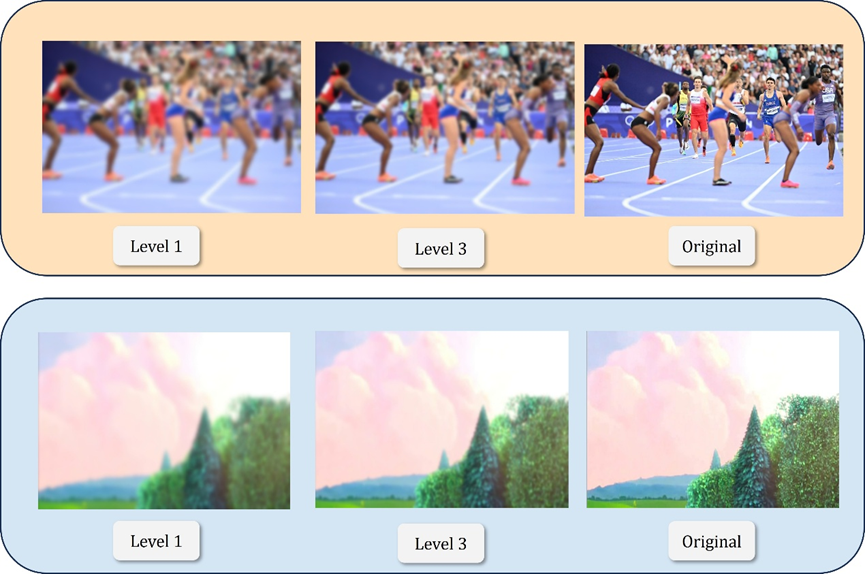}
  \caption{Same bitrate reduction, unequal semantic loss: high-motion content (top) can lose action and event-level cues, whereas static content (bottom) largely preserves high-level semantics.}
  \label{fig:scenario}
\end{figure}

Video streaming has become a major and bandwidth-intensive source of mobile data traffic in modern wireless networks. As radio resources remain limited, efficient resource allocation is critical to maintaining user viewing quality. When prioritizing and scheduling traffic, traditional wireless resource allocators focus on optimizing Quality-of-Service (QoS) metrics, such as latency, throughput, and packet loss~\cite{3gpp2018release15,stockhammer2011dash,prasad2024qos}. These metrics describe how efficiently and reliably packets traverse the network, but they do not directly capture user viewing experience.

Recognizing this gap, a growing body of work has shifted toward Quality-of-Experience (QoE)-driven resource allocation. These approaches incorporate application-layer quality signals, such as resolution, rebuffering events, and perceptual quality, into mean opinion score (MOS) prediction or optimization models~\cite{moura2023drl,ahmad2021supervised,feng2023qoe,du2023attention,liu2023qoe}. However, these signals focus on how a video looks, not what it conveys (i.e., the semantic). For example, Video Multi-Method Assessment Fusion (VMAF)~\cite{li2016vmaf} is a widely used perceptual video quality metric proposed by Netflix. VMAF and related perceptual metrics are sensitive to the level of compression. They capture visual quality rather than semantic preservation. A viewer's understanding of the video also depends on whether objects, actions, and event context survive compression. 

In adaptive bitrate streaming (ABR), each video is encoded at multiple bitrate levels, and the radio resources available to a user determine which levels can be delivered. Fig.~\ref{fig:scenario} shows that bitrate reductions can affect different videos in different ways. In an athletics race, a high-motion scene such as a sprint finish requires sufficient bitrate to preserve key semantic details. Under aggressive compression, subject roles, motion cues, action continuity, and event context may become difficult to recognize, making it harder for viewers to follow the competition and its key moments. In contrast, a static scene, such as a landscape view with minimal visual changes, can often be delivered at lower bitrates while preserving its main objects and scene context. This asymmetry, where the same bitrate reduction can lead to very different losses of meaning across videos, calls for an allocation criterion that accounts for semantic preservation rather than visual quality alone.

Multimodal large language models (MLLMs) offer a promising route for measuring such semantic effects. By jointly reasoning over visual and textual modalities, they can describe a video's semantic content in natural language~\cite{wang2024mllm}. Recent work has begun using MLLMs for video quality scoring~\cite{zhang2025qbenchvideo,cao2026vqathinker,mi2024clif}. Related semantic-aware transmission studies have used semantic information to guide resource management and video streaming~\cite{yan2024qoe,yan2025semantic}. However, in ABR, video semantics is still rarely treated as a measurable signal for bitrate selection and resource allocation. Few studies quantify how much meaning is preserved when the same video is encoded at different bitrate levels.

This paper introduces semantic fidelity (SF), a metric for measuring how much meaning is retained under compression. To compute SF, an MLLM describes different bitrate versions of a video in natural language, and a separate text-only large language model (LLM) compares the resulting descriptions. The descriptions from the MLLM cover visible subjects, actions, scene context, on-screen text, and event flow. The SF score of a compressed video is computed offline by comparing its description with that of the source reference of the same video. The resulting SF scores of compressions at different levels are cached as SF--bitrate profiles, allowing online bitrate selection to use the SF scores for resource allocation without invoking MLLMs at runtime. The profiles reveal substantial content-dependent heterogeneity: some streams require higher bitrates to preserve meaning, while others remain semantically stable at lower bitrates. This heterogeneity creates an optimization opportunity that semantic-agnostic allocation cannot exploit.

\begin{figure}[!t]
  \centering
  \includegraphics[width=\columnwidth]{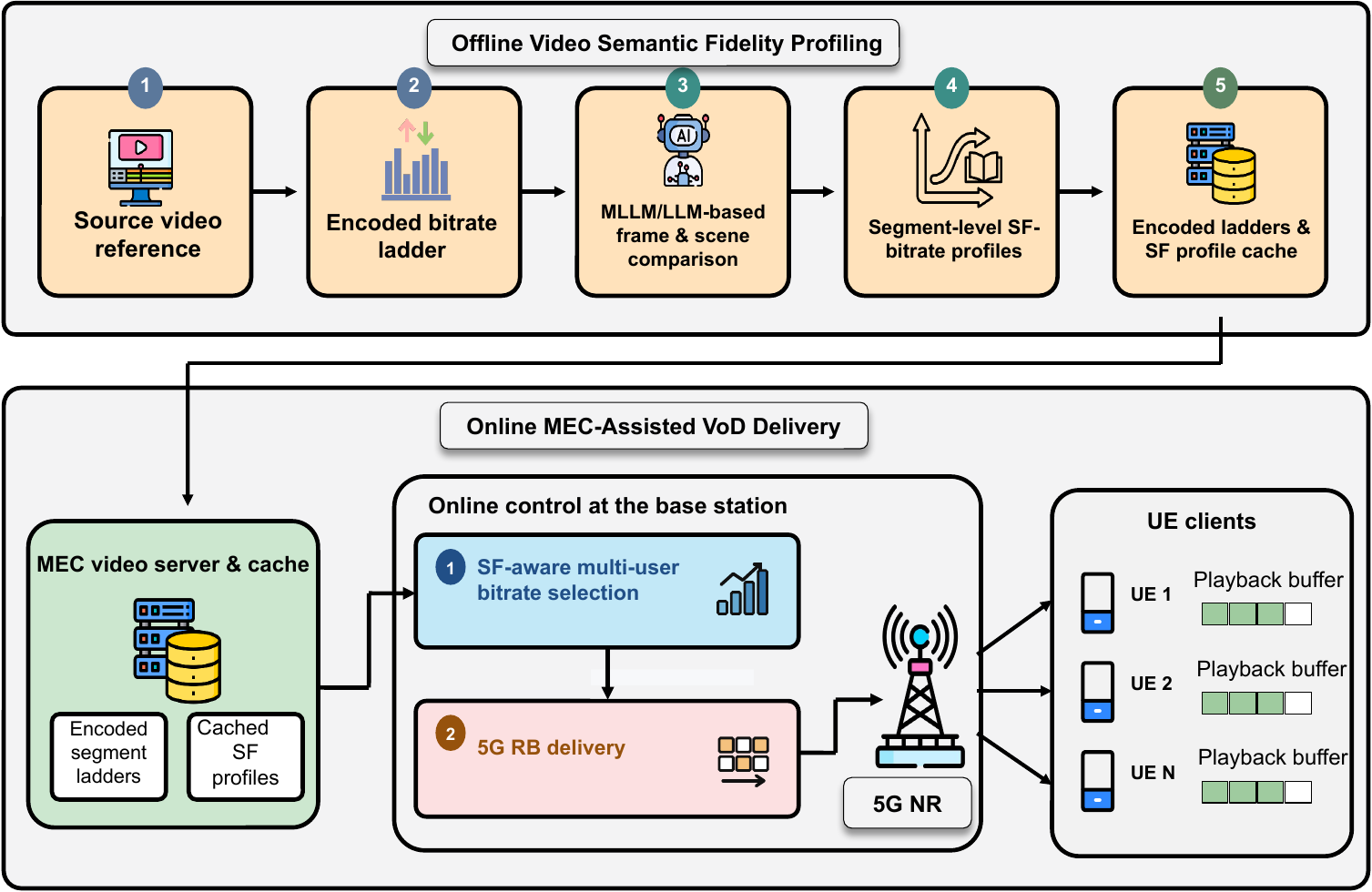}
  \caption{System overview of offline semantic-fidelity profiling and online MEC-assisted VoD delivery.}
  \label{fig:system-overview}
\end{figure}

Building on the offline SF profiles, we study a resource allocation problem in a video-on-demand (VoD) delivery system assisted by mobile edge computing (MEC). The system model consists of two parts: offline profiling and online delivery, as shown in Fig.~\ref{fig:system-overview}. For offline profiling, each video is encoded at multiple bitrate levels. The MLLM/LLM pipeline compares each bitrate version with the source reference and uses the resulting scores to construct an SF profile that relates semantic fidelity to bitrate. The MEC video server, co-located with the base station, caches these bitrate versions and their SF profiles. For online delivery, the base station allocates the available radio resources among competing video streams to optimize their semantic fidelity. The base station uses the cached SF profiles to select a bitrate level for each stream. The 5G radio scheduler then assigns radio resources to transmit each stream at its selected bitrate level.

The main contributions of this paper are as follows:

1) We define semantic fidelity (SF), an MLLM-based metric for measuring video semantic preservation. An offline pipeline compares each compressed version with the source reference of the same video and caches the resulting scores as SF--bitrate profiles across bitrate levels. Validation on three independent QoE benchmarks and a human semantic-rating study indicates that SF correlates strongly with MOS in the evaluated datasets and aligns with direct semantic-preservation judgments. In essence, SF provides a valuable resource allocation signal complementary to pixel-domain quality metrics.

2) Using the cached SF--bitrate profiles, we cast multi-user video delivery as a semantic-aware resource allocation problem in a wireless downlink. Different videos exhibit markedly different SF--bitrate relations, so that the same bitrate increase can produce very different semantic gains depending on content. The allocation framework uses this content-dependent relation to prioritize additional bitrate for semantically sensitive streams.

3) We design a two-timescale 5G resource allocation framework at the base station. At the video bitrate-selection stage, a controller maps the user playback-buffer states to a resource budget. An $\alpha$-fair multiple-choice knapsack (MCKP) formulation then selects one bitrate level for each user under this budget to reduce overall SF loss. At the wireless-scheduling stage, the base station assigns the available 5G time-frequency resource blocks to deliver the selected segments.

4) We validate the complete framework through NS-3/5G New Radio (NR) simulations across multiple system bandwidths and user loads. SF-aware allocation improves semantic fidelity over non-SF baselines, with a larger advantage when wireless resources are limited. The experiments further show that allocation without SF may not protect semantically sensitive streams sufficiently, a gap revealed by measuring semantic fidelity.

\section{MLLM-Based Semantic Fidelity Profiling for Compressed Video}
\label{sec:sf}

\begin{table*}[!t]
  \centering
  \caption{Main notation used throughout the paper.}
  \label{tab:notation}
  \footnotesize
  \setlength{\tabcolsep}{3pt}
  \renewcommand{\arraystretch}{1.08}
  \begin{tabular}{@{}>{\centering\arraybackslash}p{0.14\textwidth}
                      >{\raggedright\arraybackslash}p{0.33\textwidth}
                      >{\centering\arraybackslash}p{0.14\textwidth}
                      >{\raggedright\arraybackslash}p{0.33\textwidth}@{}}
    \toprule
    \textbf{Symbol} & \textbf{Description} & \textbf{Symbol} & \textbf{Description} \\
    \midrule
    $v$, $n$
      & Video and user indices
      & $i$, $j$, $k$, $l$
      & Section~\ref{sec:sf}: bitrate level, scene, segment, and frame; Section~\ref{sec:system}: batch index ($i$) and bitrate level ($j$) \\
    $\mathcal{BR}_v$, $q_v$, $\Delta$, $r_{v,i}$
      & Measured bitrate set, number of bitrate levels, bitrate step, and bitrate of level $i$
      & $L_v$, $F^v_{l,i}$
      & Number of frames and frame $l$ at bitrate level $i$ for video $v$ \\
    $C^v_{l,i}$, $\psi^{v,\mathrm{frame}}_{l,i}$
      & MLLM-generated caption and frame-level SF component
      & $\tau^v_{j,i}$, $\psi^v_{l,i}$
      & Scene-level temporal SF and final SF of profiling frame $l$ \\
    $w_l^v$, $W_k^v$
      & Profiling-frame and segment semantic-importance weights
      & $S_{k,i}^v$, $L_{k,i}^{v,\alpha}$
      & Segment-level SF and $\alpha$-fair semantic loss at bitrate level $i$ \\
    $CL_v^{k,i}$, $T_{\mathrm{seg}}$
      & Segment payload and physical playback duration of one segment in seconds
      & $\mathcal{N}$, $N$, $M$
      & User set, number of users, and number of resource blocks (RBs) per slot \\
    $t$, $t_i$, $T_s$
      & Slot index, starting slot of batch $i$, and slot duration in segment-display-time units
      & $B_i$, $B_{\mathrm{th}}$
      & Common buffer level in segments at the start of batch $i$ and its target \\
    $D_i$, $\hat{D}_i$, $\lambda$
      & Actual and target transmission times of batch $i$ in segment-display-time units, and buffer-smoothing parameter
      & $\eta_{n,i}$, $\hat{\eta}_{n,i}$
      & Actual and estimated RB capacities for user $n$ in batch $i$, measured in bits per RB \\
    $x_{n,j}$, $\xi$
      & Bitrate-level selection indicator and switch-penalty coefficient
      & $j_{n,i}^*$
      & Bitrate level selected for user $n$ in batch $i$ \\
    \bottomrule
  \end{tabular}
\end{table*}

\begin{figure*}[!t]
  \centering
  \includegraphics[width=\textwidth]{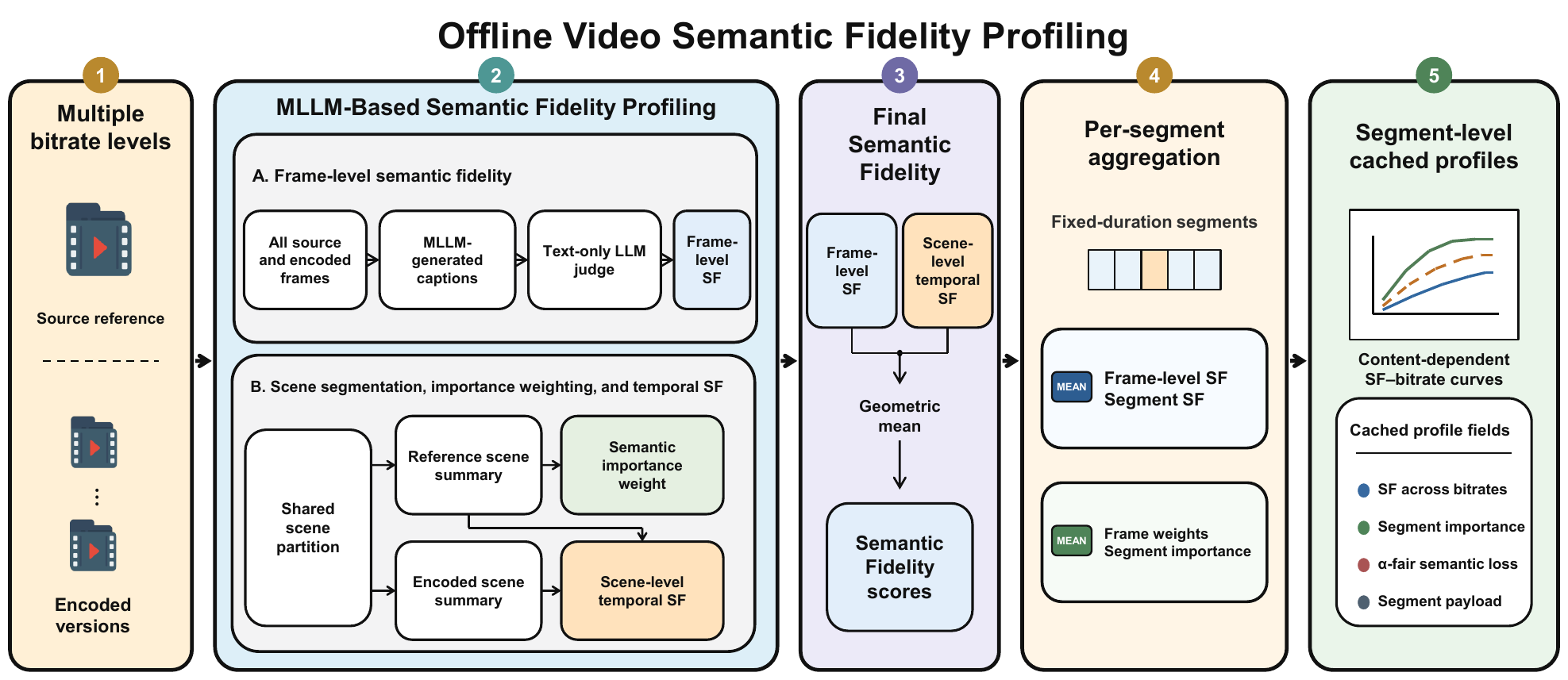}
  \caption{Detailed offline semantic-fidelity profiling pipeline. The pipeline compares MLLM-generated descriptions across bitrate versions, adds scene-level temporal and importance information, and aggregates the resulting scores into per-segment SF--bitrate profiles. The profiles are computed once offline and cached for online bitrate selection.}
  \label{fig:sf-framework}
\end{figure*}

This section describes how we compute semantic-fidelity (SF) profiles for each source video. We first construct a set of bitrate levels for each video, with the source reference at the highest level. An LLM compares the MLLM-generated caption of each compressed version with that of the source reference. The pipeline combines these comparisons with scene-level analysis and aggregates the results into segment-level profiles. Each cached profile records the SF and bit requirements associated with a segment and its available bitrate levels. Fig.~\ref{fig:sf-framework} summarizes the pipeline.

\subsection{Frame-Level Semantic Fidelity across Bitrate Levels}
\label{sec:sf-def}

We define the frame-level component of SF by comparing each compressed frame with its corresponding source-reference frame through their MLLM-generated captions. Specifically, VideoLLaMA3-7B~\cite{zhang2025videollama3} generates a structured caption for each frame, covering visible subjects, actions, scene context, and on-screen text. Qwen3-8B~\cite{qwen3technicalreport}, a separate text-only model, then serves as the semantic-similarity judge for the compressed and reference captions.

We formalize the frame-level SF computation pipeline as follows:

For each video $v$, let $r_{v,\text{max}}$ be the average bitrate of the source reference. Let $0<\Delta<r_{v,\text{max}}$ be the bitrate quantization step. The quantized target bitrates for the compressed versions, together with the source-reference bitrate, are
\begin{equation}
  \mathcal{BR}^{t}_v = \{\, \Delta,\, 2\Delta,\, \ldots,\, (q_v - 1)\Delta,\, r_{v,\text{max}} \,\},
  \quad q_v = \left\lceil \frac{r_{v,\text{max}}}{\Delta} \right\rceil,
\end{equation}
The final entry is the source reference. By construction, the final interval satisfies $0<r_{v,\text{max}}-(q_v-1)\Delta\leq\Delta$.

The first $q_v-1$ target bitrates in $\mathcal{BR}^{t}_v$ are used by the H.264/AVC encoder~\cite{jvt2005h264} to generate lower-rate versions of video $v$. The encoder output may differ slightly from the target bitrate, so the actual bitrate levels are given by the measured bitrates:
\begin{equation}
  \mathcal{BR}_v = \{\, r_{v,1},\, r_{v,2},\, \ldots,\, r_{v,q_v-1},\, r_{v,q_v} \,\},
  \label{eq:actual-br}
\end{equation}
where $r_{v,i}$ is the measured bitrate of the $i$-th lower-rate encoded version for $i=1,\ldots,q_v-1$, and $r_{v,q_v}=r_{v,\text{max}}$ is the bitrate of the source reference.

The source reference and the $q_v-1$ compressed encodings form $q_v$ bitrate-specific versions. For each video $v$ of duration $T_v$ seconds, all bitrate versions contain $L_v$ profiling frames. Frames with the same index $l$ correspond to the same temporal position across all versions of video $v$. For each version $i$, the frames form the following sequence:
\begin{equation}
  \begin{aligned}
    \mathbf{F}_{v,i}
    &= [F^v_{1,i}, F^v_{2,i}, \ldots, F^v_{L_v,i}], \\
    &\in \mathbb{R}^{L_v \times H_v \times W_v \times 3},
    \quad i=1,\ldots,q_v.
  \end{aligned}
\end{equation}
Here, $H_v$ and $W_v$ are the frame height and width, respectively, and $F^v_{l,i}$ is profiling frame $l$ of version $i$. The captioning function $\phi_{\text{caption}}$ by the MLLM maps each frame to a text description\footnote{The prompts and decoding settings for all MLLM and LLM functions are listed in Appendix~\ref{app:prompts-templates}.}:
\begin{equation}
  C^v_{l,i} = \phi_\text{caption}(F^v_{l,i}).
\end{equation}

The LLM similarity judge $\phi_{\text{similarity}}$ compares each compressed-frame caption $C^v_{l,i}$ with the source-reference caption $C^v_{l,q_v}$. It returns an integer score on a 1--9 scale. We linearly rescale this score to $[0,1]$ to define the frame-level semantic-fidelity component:
\begin{equation}
  \psi^{v,\text{frame}}_{l,i} =
  \frac{\phi_{\text{similarity}}(C^v_{l,q_v}, C^v_{l,i}) - 1}{8},
  \quad i = 1, \ldots, q_v - 1.
  \label{eq:frame-sf}
\end{equation}
The source reference is assigned perfect fidelity by definition: $\psi^{v,\text{frame}}_{l,q_v} \equiv 1$. Higher values indicate greater semantic agreement between the compressed and source-reference captions.

\subsection{Illustrative Frame-Level SF--Bitrate Behavior}
\label{sec:sf-case-study}

\begin{figure}[!t]
  \centering
  \includegraphics[width=\columnwidth]{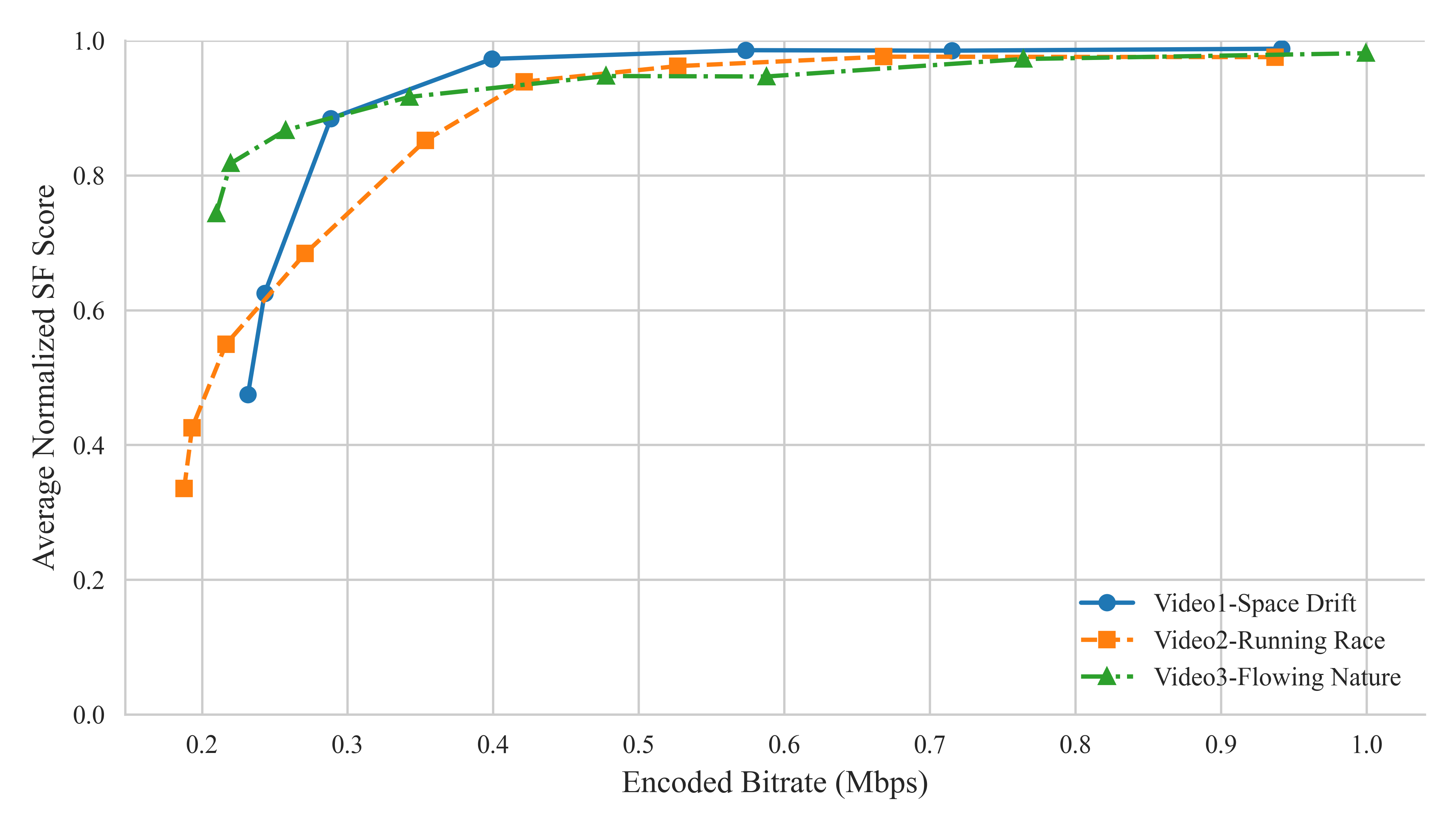}
  \caption{Average frame-level SF component $\frac{1}{L_v}\sum_{l=1}^{L_v}\psi^{v,\text{frame}}_{l,i}$ versus measured bitrate $r_{v,i}$ for the compressed encodings with $r_{v,i}\leq 1$~Mbps. Scores plateau beyond 1~Mbps, with diminishing returns toward the maximum.}
  \label{fig:sf-case-study}
\end{figure}

We evaluate three representative videos spanning a range of motion and visual complexity. \textbf{Video~1 -- Space Drift} contains relatively stable scenes with limited motion. \textbf{Video~2 -- Running Race} involves fast-moving athletes and rapid dynamic visuals. \textbf{Video~3 -- Flowing Nature} features moderate motion in natural landscapes such as rivers and forests. Fig.~\ref{fig:sf-case-study} plots the average frame-level SF against the measured bitrate for the compressed encodings with $r_{v,i}\leq 1$~Mbps for all $i$.

All three curves rise steeply at low bitrates and gradually saturate, but their rates of saturation differ. Video~3 (Flowing Nature) saturates earliest, since its natural-landscape semantics remain identifiable even under aggressive compression. Video~1 (Space Drift) follows closely: its static scenes preserve dominant objects and scene context at relatively low bitrates. Video~2 (Running Race), in contrast, requires substantially higher bitrates because its rapid dynamic visuals carry richer per-frame details that remain identifiable only at higher fidelity. Thus, the same bitrate increase does not raise SF by the same amount for different videos.

\begin{figure}[!t]
  \centering
  \includegraphics[width=\columnwidth]{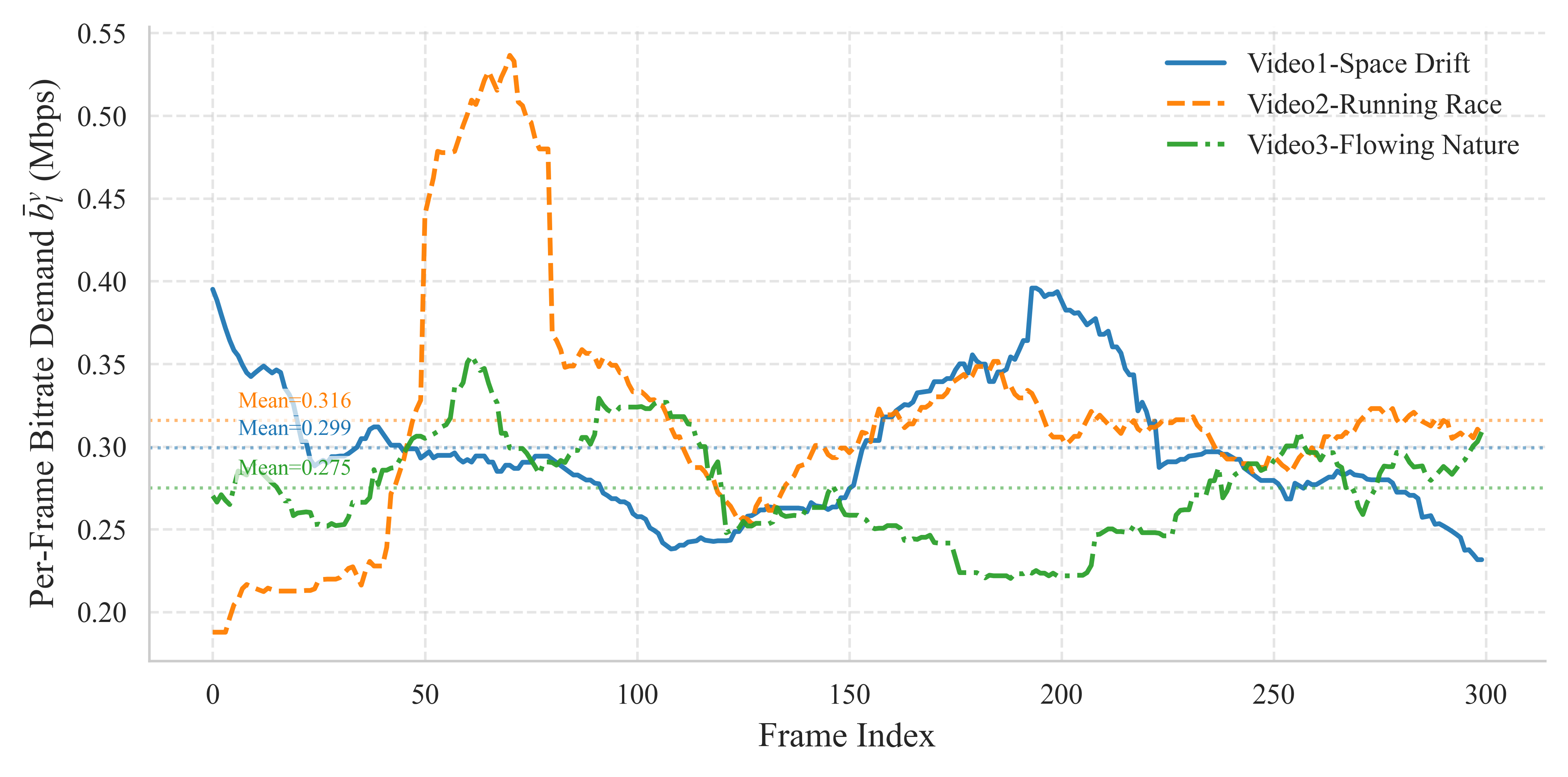}
  \caption{Per-frame bitrate demand $\bar{b}_l^v$ (1-s moving average of the minimum bitrate required to reach SF $\ge 0.875$) for the three representative videos.}
  \label{fig:sf-bitrate-demand}
\end{figure}

Beyond these inter-video differences, frame-level semantic fidelity also fluctuates substantially \emph{within} each video. We capture this with $\bar{b}_l^v$, defined as the average bitrate required to reach an SF score of 0.875 over the available profiling frames in a 1-second window centered at frame $l$:
\begin{equation}
  \resizebox{0.8\columnwidth}{!}{$\displaystyle \bar{b}_l^v = \frac{1}{|\mathcal{B}_l|}\sum_{m\in\mathcal{B}_l} \min_{r_{v,i} \in \mathcal{BR}_v}\!\{ r_{v,i} \mid \psi^{v,\text{frame}}_{m,i} \ge 0.875 \}$}.
\end{equation}
Here, $\mathcal{B}_l$ contains the profiling frames in the 1-second window centered at frame $l$. Near the beginning or end of the video, only the available frames are used. For Fig.~\ref{fig:sf-bitrate-demand}, the videos have a frame rate of 30~fps, so a full window contains 30 frames.

Fig.~\ref{fig:sf-bitrate-demand} shows that each video alternates between localized high-demand bursts (most pronounced in Video~2's running scenes) and low-demand stretches, indicating that the bitrate requirements to preserve semantics in each video are time-varying.

\subsection{Scene Segmentation and Semantic Importance Weighting}
\label{sec:sf-scene}

The frame-level SF score defined above evaluates each frame in isolation and does not account for video semantics shared across related frames. Video semantics also unfold over time, as related frames jointly convey actions and events. We treat a scene as a contiguous temporal interval that may span multiple shots but contains semantically related content. Because scene boundaries reflect the temporal organization of the source content rather than a particular compression level, we detect them from the source reference and reuse them for all compressed versions. Scenes may also differ in semantic importance, so we assign each scene a semantic-importance weight and propagate it to the scene's frames for later segment-level aggregation.

To obtain this scene partition, we use BaSSL~\cite{mun2022bassl}, a self-supervised scene-segmentation method that detects scene-level boundaries from the temporal context of neighboring shots. Applying BaSSL to the source reference produces a scene partition,
\begin{equation}
  \mathbf{Scene}_v = [\text{Scene}^v_1, \text{Scene}^v_2, \ldots, \text{Scene}^v_{J_v}],
\end{equation}
where each $\text{Scene}^v_j$ is a contiguous set of frame indices in $\{1,\ldots,L_v\}$.

For each scene, we use VideoLLaMA3-7B as the summarization function $\phi_{\text{summary}}$ to generate a scene-level caption from the scene's source-reference frames. Given the scene-level caption prompt (Appendix~\ref{app:prompts-templates}, Prompt~3), the MLLM outputs a single-paragraph caption:
\begin{equation}
  G_j^v = \phi_{\text{summary}}\!\left(\left(F^v_{l,q_v}\right)_{l \in \text{Scene}^v_j}\right),
\end{equation}
where $G_j^v$ is the scene-level caption for the $j$-th scene of video $v$.

The semantic weighting function $\phi_{\text{weight}}$ uses Qwen3-8B to assign a positive score to each scene-level caption and normalizes the scores across scenes:
\begin{equation}
  \begin{gathered}
    (W_{1,v}^{\text{Scene}},\ldots,W_{J_v,v}^{\text{Scene}})
    = \phi_{\text{weight}}(G_1^v,\ldots,G_{J_v}^v),\\
    W_{j,v}^{\text{Scene}}>0,\qquad
    \sum_{j=1}^{J_v}W_{j,v}^{\text{Scene}}=1.
  \end{gathered}
\end{equation}

Each normalized score $W_{j,v}^{\text{Scene}}$ represents the share of the video's overall semantic importance assigned to scene $j$. We distribute this scene-level weight uniformly among the scene's profiling frames and apply a common scaling factor $L_v$:
\begin{equation}
  w_{l}^{v}=\frac{L_v}{|\text{Scene}_{j}^{v}|}W_{j,v}^{\text{Scene}},
  \quad \forall l\in \text{Scene}_{j}^{v}.
\end{equation}
This definition preserves the relative scene importance and gives the frame weights unit mean, $\frac{1}{L_v}\sum_{l=1}^{L_v}w_l^v=1$, across videos of different durations.

Finally, we collect the frame-level weights as
\begin{equation}
  \mathbf{w}^{v}=[w^{v}_{1}, w^{v}_{2}, \ldots, w^{v}_{L_v}].
\end{equation}

\subsection{Temporal Semantic Fidelity via Scene-Level Narrative Comparison}
\label{sec:sf-temporal}

The frame-level score $\psi^{v,\text{frame}}_{l,i}$ defined in Section~\ref{sec:sf-def} evaluates the semantic preservation of each frame independently. It does not capture whether the temporal progression of a scene remains recognizable after compression. Compression artifacts can make the visual cues for action progression and state transitions harder to recognize. We therefore define a complementary scene-level temporal SF and combine it with the frame-level score.

Section~\ref{sec:sf-scene} has defined the source-reference scene-level caption $G_j^v$. To evaluate temporal preservation under compression, we apply the same summarization function $\phi_{\text{summary}}$ to the frames of scene $j$ in each compressed version:
\begin{equation}
  G_{j,i}^v = \phi_{\text{summary}}\!\left(\left(F^v_{l,i}\right)_{l \in \text{Scene}^v_j}\right),
  \quad i = 1, \ldots, q_v - 1,
\end{equation}
where the same scene index set $\text{Scene}^v_j$ is reused across bitrate versions while the pixel content $F^v_{l,i}$ differs. The Qwen3-8B text-only judge $\phi_{\text{temporal}}$ compares each compressed-version scene-level caption with the corresponding source-reference scene-level caption and returns a temporal-fidelity score in $[0,1]$:
\begin{equation}
  \tau^v_{j,i} = \phi_{\text{temporal}}\!\left(G_j^v,\, G_{j,i}^v\right) \in [0,1],
  \quad i = 1, \ldots, q_v - 1.
\end{equation}
The source reference is assigned perfect temporal fidelity by definition: $\tau^v_{j,q_v} \equiv 1$.

The frame-level score $\psi^{v,\text{frame}}_{l,i}$ measures semantic preservation in frame $l$, whereas the scene-level temporal score $\tau^v_{j,i}$ measures temporal preservation within scene $j$. These scores capture complementary aspects of semantic preservation, and the final SF should be high only when both aspects are preserved. For each frame index $l\in\text{Scene}_j^v$, we therefore define the final SF as their geometric mean:
\begin{equation}
  \psi^v_{l,i}
  \;\triangleq\;
  \sqrt{\psi^{v,\text{frame}}_{l,i}\tau^v_{j,i}}\in[0,1],
  \quad \forall l\in\text{Scene}_j^v,
  \quad i=1,\ldots,q_v.
  \label{eq:composite-frame-sf}
\end{equation}
The geometric mean makes the final SF sensitive to a low value in either component: a high frame-level score cannot fully compensate for a low temporal score, and vice versa. It treats the two components symmetrically and returns their common value when the scores are equal. For numerical stability, we floor each component score at $10^{-6}$ before computing the geometric mean.

\subsection{Segment-Level Cached Profiles for Adaptive Streaming}
\label{sec:sf-segment}

Adaptive streaming systems such as MPEG-DASH and HTTP Live Streaming (HLS) divide each encoded video version into segments of fixed duration $T_{\mathrm{seg}}$~\cite{stockhammer2011dash}. We group the profiling frames by segment and aggregate each group into a cached SF profile. Let $d$ be the number of frames in a full-length segment. Video $v$ contains $K_v=\lceil L_v/d\rceil$ segments. For $k=1,\ldots,K_v$, the profiling-frame index set of segment $k$ is
\begin{equation}
  \text{Seg}^v_k =
  \left\{
    l\in\{1,\ldots,L_v\}
    \,\middle|\,
    (k-1)d<l\leq kd
  \right\}.
\end{equation}
\textit{Per-segment aggregation.} Let $\boldsymbol{\psi}_l^v=[\psi_{l,1}^v,\ldots,\psi_{l,q_v}^v]$ be the SF vector of frame $l$ across bitrate levels. We aggregate these vectors using the importance weights $w_l^v$ from Section~\ref{sec:sf-scene}. The segment-level SF is the importance-weighted mean of these SF vectors, and the segment weight is the mean frame weight within the segment. The resulting segment-level values are
\begin{equation}
  \begin{aligned}
    S_{k}^{v}&=\frac{\sum\limits_{l\in \text{Seg}_{k}^{v}}w_l^v\boldsymbol{\psi}_{l}^{v}}
    {\sum\limits_{l\in \text{Seg}_{k}^{v}}w_l^v}
    \in [0,1]^{1\times q_v}, \\
    W_{k}^{v}&=\frac{1}{|\text{Seg}_{k}^{v}|}
    \sum\limits_{l\in \text{Seg}_{k}^{v}}w_{l}^{v}>0.
  \end{aligned}
  \label{eq:segment-sf}
\end{equation}
Here, $S_k^v$ is the importance-weighted SF vector of segment $k$, and $W_k^v$ is its mean semantic-importance weight. Because the segments partition the profiling frames, their weights satisfy $\frac{1}{L_v}\sum_{k=1}^{K_v}|\text{Seg}_k^v|W_k^v=1$, keeping the average weight scale independent of the video duration.

\textit{$\alpha$-fair semantic loss.} We define the segment-level $\alpha$-fair semantic loss~\cite{mo2000fair} $L_{k,i}^{v,\alpha}$ at bitrate level $i$ based on the segment's semantic importance and fidelity:
\begin{equation}
  L_{k,i}^{v,\alpha}=\begin{cases} \dfrac{W_{k}^{v}\big(1-(S_{k,i}^{v})^{1-\alpha }\big)}{1-\alpha }, & \alpha \ne 1, \\[4pt] -W_{k}^{v}\log S_{k,i}^{v}, & \alpha =1. \end{cases}
  \label{eq:alpha-loss}
\end{equation}
where $W_k^v$ and $S_{k,i}^v$ are the segment's semantic importance and SF at bitrate level $i$, respectively. The fairness parameter $\alpha\geq0$ controls the emphasis on low SF: $\alpha=0$ yields $W_k^v(1-S_{k,i}^v)$, whereas larger values penalize low-SF segments more strongly. For numerical stability when $\alpha \ge 1$, we replace $S_{k,i}^v$ in~\eqref{eq:alpha-loss} with $\max(S_{k,i}^v,\,\epsilon)$, where $\epsilon=10^{-3}$ throughout this paper.

\textit{Segment payload.} The cached profile also stores the payload $CL_v^{k,i}$, defined as the size in bits of the $k$-th encoded streaming segment at bitrate level $i$ and measured offline from the corresponding video file. It is a content-side value and is independent of the wireless channel. Thus, each segment is associated with a payload vector across all bitrate levels:
\begin{equation}
  \mathbf{CL}_{v}^{k}=[CL_{v}^{k,1},CL_{v}^{k,2},\ldots,CL_{v}^{k,q_{v}}].
\end{equation}

\section{System Model and SF-Aware Resource Allocation}
\label{sec:system}

\subsection{System Architecture and Synchronized Batch Operation}
\label{sec:system-architecture}

We consider a video-on-demand (VoD) service over a 5G network with mobile edge computing (MEC). Each encoded video is divided into segments of common playback duration $T_{\mathrm{seg}}$ seconds. A video server co-located with the base station (BS) caches each segment at multiple bitrate levels together with its pre-computed SF profile. Since the BS controls the shared downlink and can access these cached profiles, it can coordinate bitrate selection and radio-resource allocation across users.
The downlink uses 5G orthogonal frequency-division multiple access (OFDMA) and operates in transmission time slots indexed by $t\in\{0,1,2,\ldots\}$. The segment playback duration $T_{\mathrm{seg}}$ is typically much longer than the duration of a slot. Similarly, the delivery of a segment may span multiple slots. We measure time in units of $T_{\mathrm{seg}}$, so a segment has duration 1 and a slot has duration $T_s$. The BS serves the user set $\mathcal{N}=\{1,\dots,N\}$ with $M$ resource blocks (RBs) per slot. Each user is associated with one user equipment (UE).

The bitrate level of a segment remains the same during delivery, whereas RB assignments can be updated in every slot. Bitrate selection and RB scheduling therefore operate at different time scales. At the BS, we group the next segments of users into a \textbf{batch} containing $N$ segments, one per user. This grouping allows the BS to coordinate segment-level bitrate selection across users under a shared RB budget.

For $i=0,1,2,\ldots$, let $t_i$ be the slot index at which the delivery of batch $i$ begins. At $t_i$, the buffer controller sets a target RB budget for batch $i$ based on the playback-buffer state. The bitrate-level selector then selects one bitrate level for each segment in the batch in accordance with the target RB budget. Batch $i$ continues until all $N$ user segments are delivered, and the next batch begins at $t_{i+1}$. Thus, each batch spans one or more slots, and batches do not overlap. Note that the actual RB resources consumed by the batch may not be exactly equal to the target RB budget because the target RB budget was computed based on an estimation of the channel conditions, which may be different from how the channel conditions unfold during the delivery. This will manifest in the buffer settling at a level different from a target buffer level. The target RB budget for the next batch will take this into account to ensure that the buffer fluctuates only minimally around a target level.

\begin{figure}[!t]
  \centering
  \includegraphics[width=\columnwidth,height=0.45\columnwidth,keepaspectratio=false]{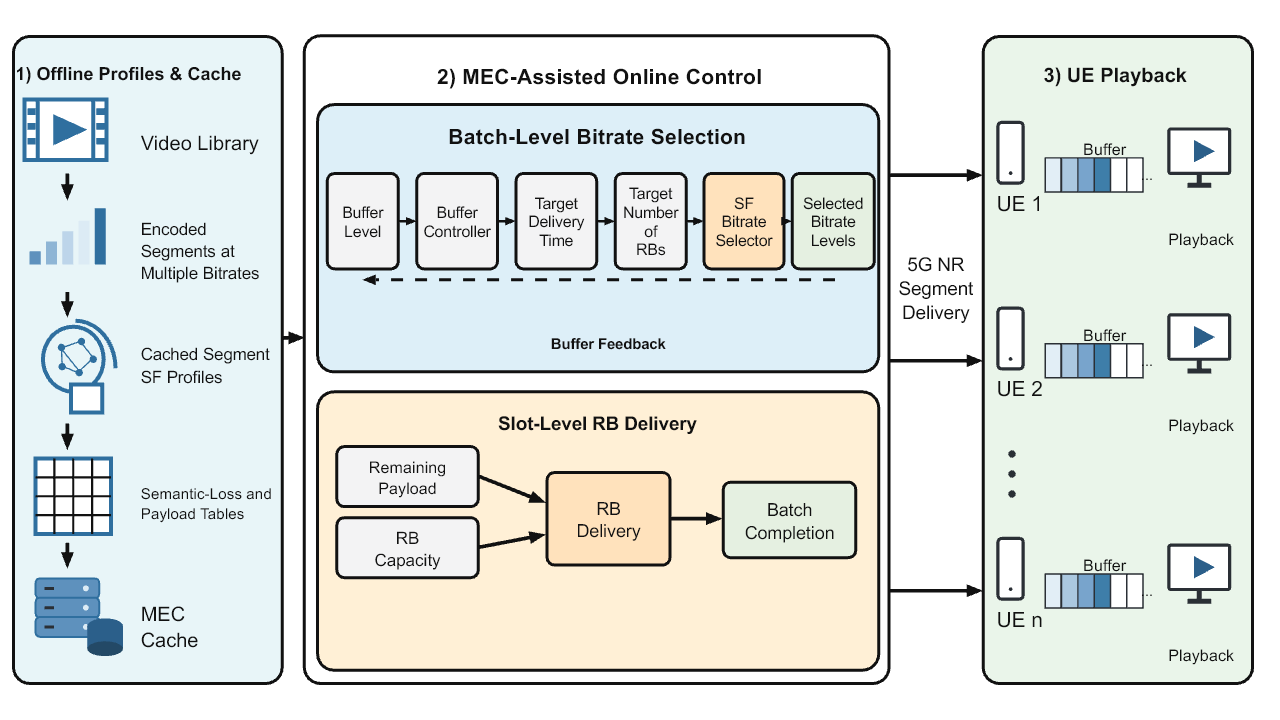}
\caption{Two-timescale SF-aware video delivery with batch-level bitrate selection and slot-level RB delivery.}
  \label{fig:system-framework}
\end{figure}

In short, the system operates over two timescales. Fig.~\ref{fig:system-framework} summarizes the two-timescale framework. The slow time scale operation is related to setting a target RB budget (and therefore the bit rates of videos) to control the buffer level and the video quality. The fast time scale is related to assigning RBs during the actual delivery of the video.

\subsection{Slow-Timescale Buffer Control and Bitrate Selection}
\label{sec:opt-formulation}

\textbf{Buffer-Based Batch Budget} At $t_0$, we assume that all users are initialized with the same number of buffered segments. Each batch contains one segment from each user, and a completed batch delivers one segment to every user's buffer. If playback does not stall, all users consume (playback) the same amount of buffered video during the delivery of a new segment because the elapsed transmission time is common to all users. Their buffer levels remain equal at each $t_i$. We can therefore use a single model to describe the buffer dynamics of all users.

Let $B_i$ be the common buffer level at $t_i$, measured in number of segments. We index the first segment delivered after initialization as segment~0. Specifically, segment~$i$ is delivered between $t_i$ and $t_{i+1}$. Thus, $B_i$ is the buffer level immediately upon the complete delivery of segment~$i-1$.

The buffer update should account for the video consumed while a batch is being transmitted, given as follows:
\begin{equation}
  B_{i+1}=B_i+1-D_i.
  \label{eq:buffer-recurrence}
\end{equation}

Between $t_i$ and $t_{i+1}$, a new segment, segment~$i$, is added to the buffer and $D_i$ video segments are consumed (i.e., have departed the buffer because of playback). The expression for the departure $D_i$ is
\begin{equation}
  D_i=(t_{i+1}-t_i)T_s.
  \label{eq:batch-duration}
\end{equation}
where $t_{i+1}-t_i$ is the number of transmission slots used to deliver the new segment and $T_s$ is the slot duration in units of a segment playback time. In other words, $D_i$ is the delivery time of segment~$i$ in unit of the duration of a segment playback.

Let $B_{\mathrm{th}}$ be a target buffer level for $B_i$. A larger target provides a greater buffer margin against variations in transmission time. The buffer controller adjusts the target transmission time of each batch to keep the buffer level close to $B_{\mathrm{th}}$. In general, the judicious setting of $B_{\mathrm{th}}$ is critical. If it is set too low, buffer underflow and freezing of video display may occur due to the random fluctuations of communication resources and video bit-rate requirements. Setting it unnecessarily high may cause the initial buffer build-up time to be longer (in general, when the user skips the video to another section, the buffer build-up time will be incurred). A model of the buffer dynamic is required to identify an optimal $B_{\mathrm{th}}$.

For the time being, let us assume that we already have a target $B_{\mathrm{th}}$. Our goal is to maintain the buffer level $B_i$ to be close to $B_{\mathrm{th}}$ by controlling the amount of time taken to deliver a segment to the buffer. The delivery time $D_i$ in \eqref{eq:batch-duration} depends on the number of bits in the new user segments and hence the quality of the video being delivered. If the current buffer level is below $B_{\mathrm{th}}$, we will need to reduce the video quality going forward, and if the current buffer level is above $B_{\mathrm{th}}$, we can afford to increase the video quality.

At $t_i$, the delivery of segment~$i-1$ has just completed, and the buffer controller can observe the buffer level $B_i$. To control the buffer level at $t_{i+1}$, $B_{i+1}$, we can focus on setting a target delivery time $\hat{D}_i$ for segment~$i$. Suppose that the controller sets $\hat{D}_i=B_i+1-B_{\mathrm{th}}$ and the actual transmission time for batch~$i$, $D_i$, turns out to be equal to this target. Then according to \eqref{eq:buffer-recurrence} and \eqref{eq:batch-duration}, $B_{i+1}$ would equal $B_{\mathrm{th}}$, as given below:
\begin{equation*}
  \begin{aligned}
    B_{i+1}
    &=B_i+1-D_i \\
    &=B_i+1-(B_i+1-B_{\mathrm{th}})
      =B_{\mathrm{th}}.
  \end{aligned}
\end{equation*}
In general, we need to bound the quantity $B_i+1-B_{\mathrm{th}}$ by $\max\{B_i+1-B_{\mathrm{th}},0\}$ since the delivery time for segment~$i$ cannot be negative. Furthermore, even if $B_i+1>B_{\mathrm{th}}$, trying to move the buffer level to $B_{\mathrm{th}}$ in one single epoch may be too aggressive and may cause large fluctuations in video quality. Before batch~0, no measured batch transmission time is available. We initialize $\hat{D}_0=1$, corresponding to one segment playback time. For $i\geq1$, we use the following smoothing scheme to smooth out the delivery time of successive segments to limit too drastic a change from one segment to the next:
\begin{equation}
  \hat{D}_i=\lambda D_{i-1}+(1-\lambda)
  \max\{B_i+1-B_{\mathrm{th}},0\},\quad i\geq1.
  \label{eq:smoothing-controller}
\end{equation}
The parameter $\lambda\in(0,1)$ controls the smoothing by setting the relative weights of the previous transmission time $D_{i-1}$ and the buffer-based term $\max\{B_i+1-B_{\mathrm{th}},0\}$. A smaller $\lambda$ gives more weight to the buffer-based term, so the target delivery time responds more strongly to the current buffer level and the buffer may return to $B_{\mathrm{th}}$ more quickly. However, the RB budget and selected bitrate levels may also change more drastically between batches. A larger $\lambda$ gives more weight to $D_{i-1}$ and keeps the target delivery time closer to the previous transmission time, producing smoother bitrate decisions but a slower buffer response. For a given $B_{\mathrm{th}}$, this slower response may increase the risk of buffer underflow.\footnote{If $B_i+1\geq B_{\mathrm{th}}$ and $D_i=\hat{D}_i$ for every batch~$i$, the buffer level converges to $B_{\mathrm{th}}$ for any $\lambda\in(0,1)$. Appendix~\ref{app:buffer-controller} provides the closed-form analysis for this case.}

\textbf{SF-Aware Bitrate Selection.} At $t_i$ (i.e., at the beginning of batch~$i$), the buffer controller first sets the target delivery time $\hat{D}_i$ based on the current buffer level. The target delivery time also determines the number of scheduling slots and hence the target number of RBs for batch~$i$. The bitrate-level selector then chooses one bitrate level for each user's next segment while keeping the total estimated RB requirement within this target. A shorter target delivery time yields a smaller number of RBs and therefore favors lower bitrate levels, whereas a larger target delivery time permits higher bitrate levels.

To select bitrate levels under the target RB budget, the bitrate-level selector estimates the RB requirement of each available bitrate level. The segment payload at each bitrate level is measured in bits, whereas the target RB budget is measured in number of RBs. Converting a segment payload into an RB requirement requires an estimate of the payload delivered per RB. Because channel conditions differ across users and change over time, the payload delivered per RB depends on both the user and batch. However, at $t_i$, the bitrate-level selector must choose the bitrate levels before the payload delivered per RB for batch~$i$ is known.

Before batch~$i$ is transmitted, the bitrate-level selector estimates the RB capacity for each user. The selector uses this capacity estimate to convert the segment payload at each bitrate level into an estimated RB requirement. We use one capacity value for each user in each batch and apply it to all RBs that may be assigned to that user. That is, we assume channel response for a particular user is flat across frequency as well as time during the batch delivery time, but the channel responses of different users may be different. Also the channel response of the same user may vary from batch to batch---i.e., slow block fading. Specifically, $\eta_{n,i}$, measured in bits per RB, is the number of payload bits that an RB can deliver to user~$n$ during batch~$i$.

Since we assume each user's RB capacity varies slowly at the batch timescale, the capacity observed in batch~$i-1$ provides a useful estimate for batch~$i$. For $i\geq1$, the selector estimates $\eta_{n,i}$ using the capacity $\eta_{n,i-1}$ observed during the delivery of user~$n$'s segment~$i-1$ in batch~$i-1$:
\begin{equation*}
  \hat{\eta}_{n,i}=\eta_{n,i-1},\qquad n\in\mathcal{N}.
\end{equation*}
For batch~0, the bitrate-level selector uses an initial estimate $\hat{\eta}_{n,0}>0$ for each user.

For each user, the number of RBs required to deliver the next segment depends on the selected bitrate level. Because the target number of RBs is limited, some bitrate-level combinations require more RBs than this target permits. The bitrate-level selector assigns one bitrate level to each user jointly according to the $\alpha$-fair semantic-loss criterion, using the estimated RB capacities $\{\hat{\eta}_{n,i}\}_{n\in\mathcal{N}}$ in the resource constraint.

For batch~$i$, let $j=1,\ldots,q_n$ index the $q_n$ bitrate levels available for user~$n$'s next segment. At bitrate level~$j$, $L_{n,j}^{\alpha}$ is the cached $\alpha$-fair semantic loss, and $CL_{n,j}$ is the corresponding segment payload in bits. Let $x_{n,j}$ be a binary variable that equals 1 when bitrate level~$j$ is selected for user~$n$, and 0 otherwise. We omit the batch index~$i$ from these symbols because the problem is formulated separately for each batch. For $i\geq1$, the bitrate-level selector solves
\begin{equation}
  \begin{aligned}
    \min_{\{x_{n,j}\}}\quad
    &\sum_{n\in\mathcal{N}}\sum_{j=1}^{q_n}
      x_{n,j}L_{n,j}^{\alpha} \\
    &+\xi\sum_{n\in\mathcal{N}}\sum_{j=1}^{q_n}
      x_{n,j}\bigl(j-j_{n,i-1}^*\bigr)^2 \\
    \text{s.t.}\quad
    &\sum_{j=1}^{q_n}x_{n,j}=1,
      \quad \forall n\in\mathcal{N}, \\
    &x_{n,j}\in\{0,1\},
      \quad \forall n\in\mathcal{N},\ j=1,\ldots,q_n, \\
    &\sum_{n\in\mathcal{N}}\sum_{j=1}^{q_n}
      x_{n,j}\left\lceil
      \frac{CL_{n,j}}{\hat{\eta}_{n,i}}
      \right\rceil
      \leq \frac{M\hat{D}_i}{T_s}.
  \end{aligned}
  \label{eq:bitrate-selection}
\end{equation}

The first term in the objective minimizes the sum of the cached $\alpha$-fair semantic losses $L_{n,j}^{\alpha}$ defined in Section~\ref{sec:sf-segment}. A larger loss indicates poorer preservation of video semantics. The second term discourages bitrate changes by imposing a penalty $\xi\bigl(j-j_{n,i-1}^*\bigr)^2$ based on the difference between the bitrate levels selected for consecutive segments. Here, $\xi\geq0$ is the switch-penalty coefficient, and $j_{n,i-1}^*$ is the bitrate level selected for user~$n$ in batch~$i-1$. The squared difference $\bigl(j-j_{n,i-1}^*\bigr)^2$ equals 0 when candidate bitrate level~$j$ is the same as the previous selection, while a larger difference produces a larger penalty. This penalty is included because selecting different levels for a user in two consecutive batches changes the bitrate between consecutive segments and may cause noticeable quality fluctuations.

The first constraint restricts that only one bitrate level is selected for each user. The second constraint defines $x_{n,j}$ as a binary variable. In the last constraint, $CL_{n,j}$ is measured in bits and $\hat{\eta}_{n,i}$ in bits per RB, so $\lceil CL_{n,j}/\hat{\eta}_{n,i}\rceil$ is the estimated number of RBs needed to deliver user~$n$'s segment at level~$j$. The target delivery time $\hat{D}_i$ corresponds to $\hat{D}_i/T_s$ slots. Since each slot contains $M$ RBs, $M\hat{D}_i/T_s$ is the target number of RBs available over the target delivery time. The last constraint therefore requires the total estimated number of RBs needed by the selected bitrate levels not to exceed this target number.

The multiple-choice knapsack problem is NP-hard in general, and a direct search would have to examine every combination of users' bitrate levels. A pseudo-polynomial dynamic programming (DP) algorithm can process the users one at a time while using the remaining number of RBs as its state. Because this state must be indexed by integers, we round each estimated RB requirement up and the target number of RBs $M\hat{D}_i/T_s$ down. Rounding in these directions makes the resource constraint stricter: any solution that satisfies the rounded constraint also satisfies the original constraint. For each of the $N$ users, the DP considers every integer state up to $\lfloor M\hat{D}_i/T_s\rfloor$ and at most $q_{\max}=\max_n q_n$ bitrate levels. The DP retains the minimum objective value for each state. Its complexity is $\mathcal{O}(Nq_{\max}\lfloor M\hat{D}_i/T_s\rfloor)$.

If even the lowest-bitrate combination requires more than $M\hat{D}_i/T_s$ RBs, \eqref{eq:bitrate-selection} has no feasible solution. The batch must nevertheless deliver one complete segment to every user. We therefore select the lowest bitrate level for all users and continue scheduling until the batch is complete. This fallback gives the smallest estimated RB requirement among the available bitrate combinations and allows the system to proceed to the next batch.

\subsection{RB Delivery and Batch Completion}

Once the bitrate levels are selected at $t_i$, the BS begins delivering the segment payloads of batch~$i$. For user~$n$, the selected bitrate level $j_{n,i}^*$ determines the payload $CL_{n,j_{n,i}^*}$. Given $M$ RBs per slot and the actual RB capacities $\eta_{n,i}$, the number of slots required to deliver all $N$ segment payloads in batch~$i$ is:
\begin{equation}
  t_{i+1}-t_i=
  \left\lceil
  \frac{1}{M}
  \sum_{n\in\mathcal{N}}
  \left\lceil
  \frac{CL_{n,j_{n,i}^*}}{\eta_{n,i}}
  \right\rceil
  \right\rceil.
  \label{eq:batch-completion}
\end{equation}
The inner ceiling gives the integer number of RBs required by each user. The sum gives the total RB requirement of batch~$i$, and the outer ceiling converts this requirement into an integer number of slots. Multiplying this slot count by the slot duration $T_s$ gives the actual delivery time $D_i=(t_{i+1}-t_i)T_s$.

The actual RB capacities $\eta_{n,i}$ may differ from the estimates $\hat{\eta}_{n,i}$ used for bitrate selection. Consequently, the actual delivery time $D_i$ may differ from its target $\hat{D}_i$. At $t_{i+1}$, the buffer controller uses $B_{i+1}$ and $D_i$ to set $\hat{D}_{i+1}$, while the observed capacity $\eta_{n,i}$ becomes the estimate $\hat{\eta}_{n,i+1}$ for the next bitrate decision. This completes one batch-level control cycle.

\section{Semantic Fidelity Evaluation on QoE Benchmarks and a Human Study}
\label{sec:validation}

This section evaluates how the proposed SF metric relates to perceived QoE and to human judgments of semantic preservation. Section~\ref{sec:validation-empirical} uses three public ABR datasets to examine the relationship between SF and the overall MOS, and whether SF provides information beyond that captured by conventional video metrics. Section~\ref{sec:validation-human} compares SF scores with direct human ratings of video semantic preservation.

\subsection{QoE Relevance and Metric Complementarity on Public ABR Datasets}
\label{sec:validation-empirical}

To evaluate SF, we use three public ABR datasets containing test videos and their mean opinion scores (MOS). Each test video is a complete ABR playback that may switch between different bitrate levels over time. Each MOS reflects viewers' overall QoE for the corresponding test video. \textbf{LIVE-NFLX-II}~\cite{bampis2021atlas} contains 420 H.264 test videos generated from 15 original videos. \textbf{Waterloo SQoE-III}~\cite{duanmu2018sqoe3} contains 450 H.264 test videos generated from 20 original videos using 6 ABR algorithms. \textbf{Waterloo SQoE-IV}~\cite{duanmu2020sqoe4} includes 225 H.264 test videos generated from 5 original videos using 45 ABR--network combinations.

Each test video is associated with an overall MOS in the dataset, and we also compute a corresponding SF score for the same test video. Let $v$ index a test video, and let $L_v$ be the number of frames in test video $v$. Following the procedure defined in Section~\ref{sec:sf}, we compute the SF $\psi_l^v$ and semantic-importance weight $w_l^v$ for each frame $l$ in test video $v$. The SF score of test video $v$ is
\begin{equation}
  \mathrm{SF}_v
  =
  \frac{\sum_{l=1}^{L_v} w_l^v\psi_l^v}
       {\sum_{l=1}^{L_v} w_l^v},
  \label{eq:video-sf}
\end{equation}

\textbf{QoE Relevance.} To compare SF with MOS, we use the Spearman rank-order correlation coefficient (SROCC) and the Pearson linear correlation coefficient (PLCC). SROCC indicates whether test videos with higher SF generally receive higher MOS, whereas PLCC indicates whether SF and MOS increase together at a consistent rate. Table~\ref{tab:table1a} reports the resulting SROCC and PLCC values for SF and the conventional video metrics across the three datasets. Across all three datasets, SF maintains SROCC above $0.80$ and PLCC above $0.75$, indicating a consistently strong association with MOS. Unlike SF, VMAF, SSIM, and PSNR do not maintain consistently strong correlations with MOS across all three datasets. We also repeat the SF evaluation using two additional MLLMs: Qwen2-VL-7B and InternVL2-8B. Both MLLMs show a similar overall pattern as VideoLLaMA3-7B, with SF consistently associated with MOS across all three datasets (Appendix~\ref{app:cross-mllm}).

\begin{table}[!t]
  \centering
  \caption{SROCC and PLCC of SF and conventional video metrics against human MOS.}
  \label{tab:table1a}
  \scriptsize
  \setlength{\tabcolsep}{3pt}
  \begin{tabular*}{\columnwidth}{@{\extracolsep{\fill}}l cc cc cc@{}}
    \toprule
    \textbf{Metric} & \multicolumn{2}{c}{\textbf{LIVE-NFLX-II}} & \multicolumn{2}{c}{\textbf{SQoE-III}} & \multicolumn{2}{c}{\textbf{SQoE-IV}} \\
    \cmidrule(lr){2-3} \cmidrule(lr){4-5} \cmidrule(lr){6-7}
    & SROCC & PLCC & SROCC & PLCC & SROCC & PLCC \\
    \midrule
    \textbf{SF} & 0.8096 & 0.7573 & \textbf{0.8104} & \textbf{0.7799} & \textbf{0.8389} & \textbf{0.8705} \\
    VMAF & \textbf{0.8267} & \textbf{0.8507} & 0.5613 & 0.6723 & 0.7856 & 0.8553 \\
    ST-RRED & 0.7171 & 0.7188 & 0.4706 & 0.2553 & 0.6324 & 0.4485 \\
    MS-SSIM & 0.6954 & 0.7210 & 0.5214 & 0.5895 & 0.5713 & 0.4293 \\
    SSIM & 0.7036 & 0.7302 & 0.5246 & 0.3527 & 0.7149 & 0.7478 \\
    PSNR & 0.6688 & 0.6812 & 0.4606 & 0.4953 & 0.7145 & 0.7015 \\
    \bottomrule
  \end{tabular*}
  \\[3pt]
  {\scriptsize Note: Because ST-RRED~\cite{soundararajan2013strred} uses the opposite scoring direction, we reverse its correlation signs for comparison with the other metrics.}
\end{table}

\begin{figure}[!t]
  \centering
  \includegraphics[width=0.75\columnwidth]{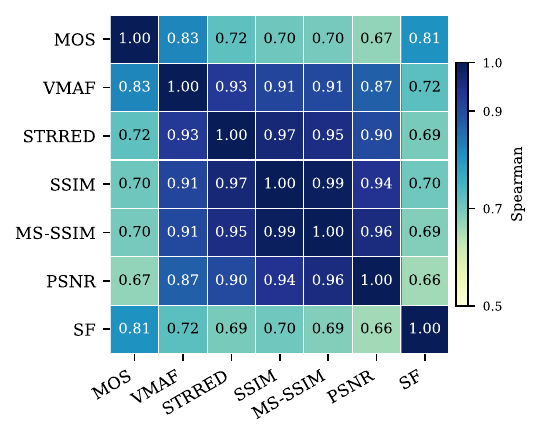}
  \caption{Pairwise SROCC matrix on LIVE-NFLX-II.}
  \label{fig:pairwise-srocc}
\end{figure}

\textbf{Comparison with Video Metrics.} Fig.~\ref{fig:pairwise-srocc} shows the pairwise SROCC between SF, conventional video metrics, and MOS on LIVE-NFLX-II. Correlations among the metrics indicate how similarly they score the test videos, whereas correlations with MOS indicate how closely each metric follows viewers' overall ratings. As shown in Fig.~\ref{fig:pairwise-srocc}, the five conventional video metrics have pairwise SROCC values of at least $0.87$, showing that, to a large extent, they score videos similarly. SF has lower SROCC values of $0.66$--$0.72$ with these metrics but retains the second-highest SROCC with MOS. This pattern indicates that SF does not simply reproduce the scores of conventional video metrics. VMAF achieves a higher SROCC of $0.83$ on this dataset, but the strong correlation between SF and MOS suggests that semantic preservation is also relevant to viewers' overall QoE and warrants greater attention in video-streaming research.

\subsection{Human Evaluation of Semantic Preservation}
\label{sec:validation-human}

Section~\ref{sec:validation-empirical} shows a positive association between SF and overall MOS. In the MOS datasets, viewers rated their overall viewing experience rather than semantic preservation specifically -- i.e., they were not explicitly asked to focus on semantic preservation when the dataset was collected. Our main result was that SF, being a semantic preservation metric, is positively associated with quality of experience as measured by MOS. This implies that human viewers may have subconsciously and indirectly factored semantic preservation into their quality of experience.

In this section, we examine whether SF agrees with direct human judgments of semantic preservation. Specifically, unlike the MOS dataset in Section~\ref{sec:validation-empirical}, for the dataset here, participants were explicitly asked to rate how well each compressed video preserved the semantic content of its source reference video.

We selected $30$ source-reference videos from LIVE-NFLX-II and SQoE-III and encoded each video at four bitrates ($200$~kbps, $500$~kbps, $1.5$~Mbps, and $4$~Mbps), producing $120$ compressed videos. Each compressed video was paired with its source reference for evaluation.

Participants viewed each compressed video and its source reference side by side, with their left--right positions randomized. Participants then rated how well each compressed video preserved the semantic content of its source reference on a $1$--$9$ scale, with higher scores indicating better semantic preservation. We averaged the ratings for each compressed video.

\begin{figure}[!t]
  \centering
  \includegraphics[width=\columnwidth]{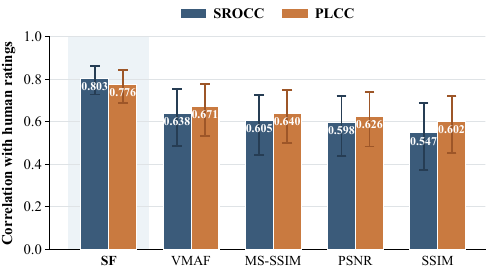}
  \caption{SROCC and PLCC between each metric and human ratings of semantic preservation.}
  \label{fig:human-ratings}
\end{figure}

Fig.~\ref{fig:human-ratings} shows the SROCC and PLCC between each evaluated metric and the average human rating. SF has the highest SROCC and PLCC among all evaluated metrics. Among the four conventional video metrics, VMAF has the highest correlations. SF exceeds VMAF by $0.17$ in SROCC and $0.11$ in PLCC. The higher SROCC and PLCC values indicate that SF is more closely associated with human ratings of semantic preservation than the conventional video metrics.

In the human study, SF correlates more strongly with human ratings of semantic preservation than any of the evaluated conventional video metrics. SF also maintains a consistent association with MOS across the three datasets. Taken together, these findings indicate that semantic preservation is a relevant component of streaming QoE but is not fully captured by conventional video metrics. We therefore argue that semantic preservation should be considered a distinct component of streaming QoE, alongside visual quality and playback factors.

\FloatBarrier

\section{System-Level Evaluation}
\label{sec:experiments}

\subsection{Simulation Setup}
\label{sec:sim-framework}

We implement the SF-aware framework described in Section~\ref{sec:system} using the 5G NR module in NS-3. SF-Aware uses the buffer controller in Section~\ref{sec:system} to adjust the target RB budget for each batch and keep the common buffer level close to $B_{\mathrm{th}}$. The simulations consider a single-cell downlink in which one BS serves all UEs. We vary the number of UEs and system bandwidth across the experiments. Each UE receives a different video so that the evaluation covers different content-dependent SF--bitrate profiles. For each video, we construct the available bitrate levels and cached segment-level SF--bitrate profiles as described in Section~\ref{sec:sf}. Table~\ref{tab:sim-params} summarizes the simulation parameters.

\begin{table}[!b]
  \centering
  \footnotesize
  \caption{Key simulation parameters.}
  \label{tab:sim-params}
  \setlength{\tabcolsep}{3pt}
  \begin{tabular*}{\columnwidth}{@{\extracolsep{\fill}}l l@{}}
    \toprule
    \textbf{Parameter} & \textbf{Value} \\
    \midrule
    NS-3 version & 3.39 + NR module \\
    Carrier frequency & 4~GHz \\
    Numerology & 1 (30~kHz subcarrier spacing) \\
    Default scenario & 3GPP UMi Street Canyon \\
    BS transmit power & 43~dBm \\
    System bandwidths & 15 / 12 / 10~MHz \\
    Segment duration $T_{\mathrm{seg}}$ & 2.0~s  \\
    Bitrate step $\Delta$ & 500~kbps \\
    Target buffer level $B_{\mathrm{th}}$ & 5 segments \\
    Buffer-smoothing parameter $\lambda$ & 0.5 \\
    Fairness parameter $\alpha$ & 1 \\
    Switch-penalty coefficient $\xi$ & 0.4 \\
    \bottomrule
  \end{tabular*}
\end{table}

\subsection{Comparison with Baselines}
\label{sec:baseline}

We compare SF-Aware with six baselines in three groups: 1) conventional ABR algorithms, 2) equal resource sharing and 3) variants that keep the same bitrate-level selector but replace SF with other scoring signals.

The two conventional ABR baselines, BOLA~\cite{spiteri2020bola} and RobustMPC~\cite{yin2015mpc}, do not use the buffer controller proposed in Section~\ref{sec:system}. Instead, each video stream has a separate playback buffer located at its UE, and the corresponding ABR algorithm selects the bitrate of the next segment. For both baselines, the BS uses the QoS-aware scheduler in the NS-3 5G NR module~\cite{koutlia2023qos} to allocate RBs for the requested segments. \textbf{BOLA} independently selects the bitrate of each requested segment at each UE based on the local buffer level and the utility and payload of each available bitrate level. \textbf{RobustMPC} selects the next bitrate by balancing video quality, bitrate changes, and rebuffering risk over a future horizon. Together, BOLA and RobustMPC serve as representative baselines for two common classes of ABR algorithms: buffer-based methods and model predictive control (MPC)-based ABR methods, respectively.

Unlike BOLA and RobustMPC, the four remaining baselines use the same BS-side buffer controller as SF-Aware to determine the total target RB budget for each batch. \textbf{Equal-Share Budget (ES)} divides this total RB budget equally among the UEs. For each UE, ES selects the highest bitrate level whose estimated RB requirement does not exceed its assigned share. The other three baselines use the same bitrate-level selector as SF-Aware but replace the SF scores with scores from other metrics. \textbf{VMAF-Aware} uses perceptual video-quality scores from VMAF~\cite{li2016vmaf}. \textbf{P.1203-Aware} uses scores from P.1203~\cite{robitza2017p1203}, a standardized QoE model that combines video quality with playback events such as stalling and quality changes to produce an overall QoE score. \textbf{CLIP-Aware} uses CLIP similarity scores~\cite{radford2021clip}. These baselines compare SF with perceptual video quality, QoE, and CLIP similarity under the same bitrate-level selector.

For each UE $n$, let $S_k$ be the cached SF of its delivered segment $k$ at the selected bitrate level. Let $W_k$ be the segment's semantic-importance weight defined in Section~\ref{sec:sf-segment}. We compute the overall SF $\bar{S}_n$ of the stream delivered to UE $n$ over the same segment range for all schemes:
\begin{equation}
  \bar{S}_n = \frac{\sum_k W_k S_k}{\sum_k W_k}.
  \label{eq:delivered-video-sf}
\end{equation}

\begin{table}[!t]
  \centering

  \caption{Baseline comparison using SF at $N=6$ and 10~MHz. Values are averages over ten simulation runs with different random seeds.}
  \label{tab:baseline-n6-new}
  \footnotesize
  \setlength{\tabcolsep}{2pt}
  \renewcommand{\arraystretch}{1.05}
  \begin{threeparttable}
  \begin{tabular*}{\columnwidth}{@{\extracolsep{\fill}}l c c c c c@{}}
    \toprule
    Scheme & \shortstack{Avg.\\SF} & \shortstack{Worst-User\\SF} & \shortstack{Sw./min} & \shortstack{Stalls/\\UE} & \shortstack{Jain's\\Index} \\
    \midrule
    \textbf{SF-Aware} & \textbf{0.931} & \textbf{0.908} & \textbf{4.00} & 0.00 & \textbf{0.9997} \\
    ES                            & 0.859 & 0.699 & 20.00 & 0.00 & 0.9895 \\
    BOLA~\cite{spiteri2020bola}   & 0.856 & 0.683 & 24.50 & 0.00 & 0.9881 \\
    RobustMPC~\cite{yin2015mpc}   & 0.869 & 0.705 & 18.25 & 0.33 & 0.9905 \\
    VMAF-Aware~\cite{li2016vmaf}  & 0.907 & 0.787 & 5.50 & 0.00 & 0.9963 \\
    P.1203-Aware~\cite{robitza2017p1203} & 0.893 & 0.790 & 4.75 & 0.00 & 0.9945 \\
    CLIP-Aware~\cite{radford2021clip}    & 0.853 & 0.680 & \textbf{4.00} & 0.00 & 0.9830 \\
    \bottomrule
  \end{tabular*}
  \begin{tablenotes}[flushleft]
    \footnotesize
    \item[] \emph{Note:} Avg. SF is the mean of $\bar{S}_n$ across UEs, whereas Worst-User SF is the minimum $\bar{S}_n$ among UEs. Jain's index is $J=(\sum_{n=1}^{N}\bar{S}_n)^2/(N\sum_{n=1}^{N}\bar{S}_n^2)$. Switching frequency is the average number of bitrate-level changes per UE per minute of playback. Stalls/UE is the number of playback stalls caused by buffer depletion, averaged over UEs.
  \end{tablenotes}
  \end{threeparttable}
\end{table}

Table~\ref{tab:baseline-n6-new} shows that SF-Aware attains the highest average and worst-user SF. Compared with the strongest non-SF result for each metric, SF-Aware increases average SF by 0.024 and worst-user SF by 0.118. All schemes using the BS-side buffer controller complete without stalls, whereas RobustMPC averages 0.33 stalls per UE. BOLA also avoids stalls, but this comes at the cost of the highest switching frequency, averaging 24.50 switches per minute. Overall, SF-Aware improves semantic fidelity, particularly for the worst user, while maintaining a low switching frequency.

\subsection{SF Evaluation of Delivered Videos Using Different Models}
\label{sec:independent-e2e}

The comparison in Table~\ref{tab:baseline-n6-new} uses the cached SF profiles that also guide bitrate selection in SF-Aware. Since SF-Aware is designed to improve the SF scores in these profiles, evaluating its performance with the same scores may favor SF-Aware. We therefore also evaluate the delivered videos using a different MLLM and a different LLM judge. That is, the cached SF profiles used for resource allocation are obtained based on an MLLM and an LLM judge. But for evaluation, we use a different MLLM and a different LLM judge.

Specifically, we use the SF pipeline in Section~\ref{sec:sf} to generate a new set of SF--bitrate profiles using InternVideo2.5-8B~\cite{wang2025internvideo25} and Prometheus~2 7B~\cite{kim2024prometheus2} in place of VideoLLaMA3-7B and Qwen3-8B, respectively. For each scheme, we then evaluate its SF using the new profiles according to~\eqref{eq:delivered-video-sf}, keeping its bitrate selections unchanged. We also report VMAF and P.1203 for the delivered videos to compare their perceptual-quality and QoE results.

\begin{figure}[!t]
  \centering
  \includegraphics[width=\columnwidth]{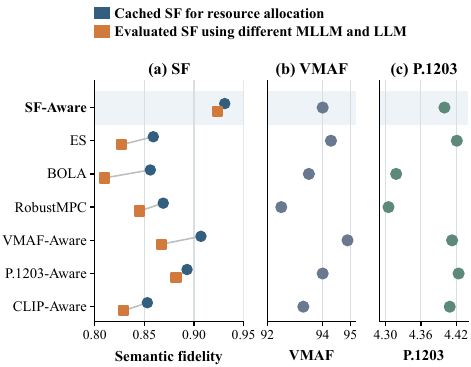}
  \caption{SF, VMAF, and P.1203 for SF-Aware and six baselines at $N=6$ and 10~MHz. In panel (a), the blue markers use cached SF from VideoLLaMA3-7B and Qwen3-8B, and the orange markers show SF evaluated using InternVideo2.5-8B and Prometheus~2 7B.}
  \label{fig:independent-e2e}
\end{figure}

Fig.~\ref{fig:independent-e2e} reports SF from both sets of models, together with VMAF and P.1203, for all seven schemes. SF-Aware achieves the highest SF with both sets of models, showing that its advantage over the evaluated baselines is preserved when a different MLLM and LLM judge are used for evaluation. Future work could also integrate SF into a complete QoE model or maximize SF subject to explicit constraints based on perceptual-quality metrics and a standardized streaming-QoE model.

\subsection{Load and Bandwidth Effects}
\label{sec:scalability}

When the available wireless resources cannot support high bitrate levels for all UEs, the bitrate-selection rule determines which streams receive higher or lower bitrate levels. The selected bitrate levels affect how much semantic content is preserved in each UE's video. We examine tighter resource constraints in two ways: by increasing the number of UEs at a fixed bandwidth and by reducing the system bandwidth for a fixed number of UEs.

\begin{figure}[!t]
  \centering
  \includegraphics[width=\columnwidth]{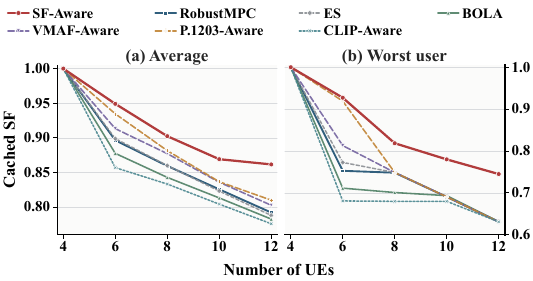}
  \caption{SF versus the number of UEs at 15~MHz: (a) average and (b) worst-user values.}
  \label{fig:scalability-load}
\end{figure}

Fig.~\ref{fig:scalability-load} examines the effect of user load by varying user $N$ from 4 to 12 at a fixed bandwidth of 15~MHz. At $N=4$, both average and worst-user SF reach the maximum value of 1.0 for every scheme. The schemes cannot be distinguished in this low-load setting because the resource constraint does not force a reduction in SF. As $N$ increases, both average and worst-user SF decrease. SF-Aware achieves the highest values for $N=6$--$12$, and its advantage is most pronounced for worst-user SF at larger $N$. The SF--bitrate profiles capture how the semantic fidelity of each video changes across bitrate levels, while the $\alpha$-fair semantic loss places greater emphasis on avoiding low-SF outcomes. Together, these components help protect users whose semantic fidelity would otherwise degrade most as resource competition increases.

\begin{figure}[!t]
  \centering
  \includegraphics[width=\columnwidth]{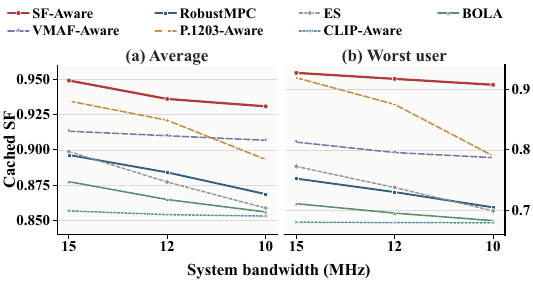}
  \caption{SF versus system bandwidth at $N=6$: (a) average and (b) worst-user values.}
  \label{fig:bandwidth-comparison}
\end{figure}

Fig.~\ref{fig:bandwidth-comparison} shows the effect of system bandwidth when the number of UEs is fixed at $N=6$. As the bandwidth decreases from 15 to 10~MHz, both average and worst-user SF decrease overall. SF-Aware achieves the highest values at every bandwidth, and its margin over the strongest baseline widens as the bandwidth decreases. The larger gains at lower bandwidths indicate that SF provides more useful guidance for bitrate selection when wireless resources are limited.

\section{Discussion and Conclusion}
\label{sec:conclusion}

This paper investigated how video semantic preservation can be measured and used when limited wireless resources force bitrate tradeoffs among video streams. We introduced video semantic fidelity (SF), a metric for quantifying the semantic content retained by a compressed video relative to its source reference. We developed an offline MLLM/LLM pipeline to construct content-specific SF--bitrate profiles for adaptive video delivery. The resulting profiles show that the same bitrate reduction can produce different semantic losses across videos.

We evaluated whether SF is positively associated with overall QoE and aligns with human judgments of semantic preservation. Across three public ABR datasets, SF shows a consistent positive association with MOS. In the human study, SF corresponds more closely to direct judgments of semantic preservation than the evaluated conventional video metrics. We therefore argue that video semantic fidelity should be considered a distinct component of streaming QoE, alongside visual quality and playback factors.

We further evaluated SF as an input signal for resource allocation. We incorporated the SF--bitrate profiles into a 5G MEC-assisted VoD framework, allowing bitrate selection to account for differences in semantic loss across videos. SF-Aware matches the baselines in the saturated low-load setting and achieves higher average and worst-user SF as the number of users increases or the system bandwidth decreases. SF-Aware also retains the highest SF when the delivered videos are evaluated using a different MLLM and a different LLM judge. These results indicate that SF provides useful guidance for bitrate selection when limited wireless resources require tradeoffs among video streams.

The current study has several limitations. First, the SF pipeline is based on MLLM-generated descriptions and LLM judgments, which may not capture all fine-grained visual features. Second, the evaluation focuses on the association between SF and overall QoE rather than its integration into a complete QoE model. Finally, the resource-allocation experiments consider synchronized batch operation; extending the framework to asynchronous delivery remains future work.

\par\medskip\noindent\textbf{Future research directions.}
The results motivate three directions for future work:
\begin{enumerate}[label=\arabic*),leftmargin=*,itemsep=2pt,topsep=2pt]
  \item \textbf{Larger semantic-fidelity benchmarks.} A public benchmark with direct human ratings would provide a common basis for developing and comparing semantic-fidelity metrics. Such a benchmark should cover diverse content, compression levels, codecs, languages, and temporal structures. Ratings of retained semantic content should be collected separately from visual-quality and overall-QoE ratings.
  \item \textbf{More accurate and efficient SF computation.} Future work should improve the recognition of identities and specialized content while reducing the computational cost of offline profiling. Smaller MLLMs and selective frame processing are possible directions for reducing this cost. Image features or visual-token similarity could also be incorporated into SF computation to capture fine-grained visual details.
  \item \textbf{Broader integration into adaptive delivery.} Future systems could combine SF with visual quality and playback factors in a complete QoE model. Evaluating the resulting framework under asynchronous delivery, rebuffering, and a wider range of network conditions would extend the current system-level evaluation.
\end{enumerate}

\appendices
\section{Prompt Templates, Profiling Cost, and Additional SF Validation}
\label{app:prompts}

\setcounter{table}{0}
\renewcommand{\thetable}{A.\Roman{table}}

Sections~\ref{sec:sf-def}, \ref{sec:sf-scene}, and~\ref{sec:sf-temporal} define five model-based functions ($\phi_{\text{caption}}$, $\phi_{\text{similarity}}$, $\phi_{\text{summary}}$, $\phi_{\text{weight}}$, $\phi_{\text{temporal}}$). Together, these functions produce the Semantic Fidelity score and the scene-level importance weights. Section~\ref{app:prompts-templates} provides representative prompt templates for each function. Section~\ref{app:cross-mllm} provides additional validation of the SF formulation. All prompts are invoked with greedy decoding (\texttt{temperature = 0}, \texttt{top\_p = 1}).

\subsection{Prompt templates}
\label{app:prompts-templates}

\smallskip
\noindent\textit{Prompt 1 --- Frame caption ($\phi_{\text{caption}}$).} Input: a single decoded video frame. Output: a structured description with four labeled semantic fields followed by a one-sentence summary that condenses all present fields.
\begin{promptbox}\scriptsize\ttfamily
You are an objective video-frame captioner. Analyze the provided frame and produce exactly five labeled lines in the order below. Omit content for any field that is genuinely absent from the frame; do not invent details. Do not mention image quality, compression, or production aspects.\\[3pt]
SUBJECT: Identify the primary subject (the most salient person, animal, or object). Include count, gender or species, distinguishing clothing or colors, and body pose.\\[2pt]
ACTION: Describe, in one sentence, the action or physical state the subject is currently performing (e.g., sprinting toward the net, standing at a podium, lying on a sofa).\\[2pt]
CONTEXT: Describe the scene setting in one sentence, covering location or environment, time of day, weather conditions, and the most salient background elements.\\[2pt]
ON-SCREEN TEXT: Transcribe verbatim any clearly readable text visible in the frame (signs, captions, overlays, scoreboards, HUD elements, jersey numbers). Write NONE if no text is present.\\[2pt]
SUMMARY: Write one self-contained sentence that integrates all the above present fields into a coherent description of the frame.\\[3pt]
Output exactly these five labeled lines and nothing else.
\end{promptbox}

\smallskip
\noindent\textit{Prompt 2 --- Caption-pair similarity ($\phi_{\text{similarity}}$).} Input: a reference structured caption and a candidate structured caption, both produced by Prompt~1 for the same frame. Output: an integer score in $\{1,2,\ldots,9\}$.
\begin{promptbox}\scriptsize\ttfamily
You are an objective evaluator of semantic equivalence between two structured video-frame captions. Each caption contains the labeled fields SUBJECT, ACTION, CONTEXT, ON-SCREEN TEXT, and SUMMARY, produced for the same frame under different compression levels. Score how much of the REFERENCE caption's semantic content is preserved in the CANDIDATE on the following 1-to-9 integer scale. Evaluate each field independently, then derive a holistic score:\\[3pt]
9 -- Identical across all fields. Same subject identity, action, context, and on-screen text (if any).\\[2pt]
7 -- Substantially equivalent. Minor descriptive details differ in at most one field; the main subject, action, and context agree.\\[2pt]
5 -- Partially equivalent. Subject matches but at least one of ACTION, CONTEXT, or ON-SCREEN TEXT is wrong, vague, or missing.\\[2pt]
3 -- Largely different. Only superficial overlap across fields; a reader of the CANDIDATE would misidentify the frame's content.\\[2pt]
1 -- Completely unrelated.\\[3pt]
If multiple fields degrade simultaneously, choose the lower candidate score (e.g., SUBJECT correct but both ACTION and CONTEXT wrong $\Rightarrow$ score~3). ON-SCREEN TEXT mismatches are weighted equally to ACTION mismatches. Output the integer score and nothing else.\\[3pt]
REFERENCE: \{C\_ref\}\\
CANDIDATE: \{C\_cand\}
\end{promptbox}

\smallskip
\noindent\textit{Prompt 3 --- Scene summarization ($\phi_{\text{summary}}$).} Input: the ordered sequence of frames belonging to a single scene cluster $\text{Scene}_j^v$. Output: a one-paragraph scene-level caption.
\begin{promptbox}\scriptsize\ttfamily
You are an objective video scene summarizer. The N frames provided are temporally ordered profiling frames from a single scene of a video. Describe the scene in one paragraph of three to five sentences. The summary should cover (i) the main subjects and their identifying attributes, (ii) the events or activities that occur within the scene, (iii) the environment or setting, and (iv) how the events or states progress within the scene. Do not mention frame numbers, image quality, compression, or production aspects. Output the summary paragraph and nothing else.
\end{promptbox}

\smallskip
\noindent\textit{Prompt 4 --- Scene-importance weighting ($\phi_{\text{weight}}$).} Input: the ordered list of $J_v$ scene-level captions $(G_1^v, \ldots, G_{J_v}^v)$. Output: $J_v$ real numbers in $(0,1)$, each reflecting the corresponding scene's relative contribution to the overall video narrative. These scores are normalized across scenes before use.
\begin{promptbox}\scriptsize\ttfamily
You are an objective evaluator of scene-level semantic importance. You will receive J textual summaries describing the consecutive scenes of a single video, in playback order. Read all J summaries together to understand the video's overall narrative arc, then assign each scene a strictly positive importance score in (0,1) reflecting its relative contribution to the viewer's understanding of the video as a whole. Higher scores indicate scenes that carry distinctive subjects, pivotal events, or content that the viewer would be unlikely to recover from neighboring scenes. Lower scores indicate scenes that are repetitive, transitional, or whose content is largely redundant with adjacent scenes. Do not assign zero to any scene. The J scores need not sum to a fixed value. Output the J scores as a comma-separated list in playback order and nothing else.\\[3pt]
SCENE 1: \{summary\_1\}\\
SCENE 2: \{summary\_2\}\\
\ldots\\
SCENE J: \{summary\_J\}
\end{promptbox}

\smallskip
\noindent\textit{Prompt 5 --- Temporal-narrative judge ($\phi_{\text{temporal}}$).} Input: a pair of scene narratives produced by $\phi_{\text{summary}}$ on the same scene, the first from the source reference and the second from a compressed version. Output: a JSON object containing a temporal-fidelity score in $[0,1]$.
\begin{promptbox}\scriptsize\ttfamily
You are an objective evaluator of how well a compressed video preserves the temporal narrative of a reference video at the scene level. You see only two textual scene descriptions, not the videos themselves. Internally weigh, in the following order, (i) subject and identity persistence, (ii) action progression, (iii) event order, (iv) state transitions of subjects and objects, (v) interactions between subjects and objects, (vi) missing short but pivotal events, and (vii) hallucinated or contradictory events. If multiple aspects degrade simultaneously, choose the lower of the candidate scores. Do not output sub-scores, reasoning, or natural-language explanations. Output exactly one JSON line and nothing else, of the form \{"temporal\_semantic\_fidelity": x\}, where x is a real number in [0,1].\\[3pt]
REFERENCE: \{C\_ref\}\\
CANDIDATE: \{C\_cand\}
\end{promptbox}

\FloatBarrier

\subsection{Offline profiling cost}
\label{app:sf-cost}

The SF pipeline runs once per source video, and the cached profiles are reused during online allocation. Table~\ref{tab:sf-cost} reports representative per-unit computation times measured on a single NVIDIA RTX~3080 GPU (20\,GB). The total profiling cost depends on the number of decoded frames, bitrate levels, and detected scenes in each source video.

\begin{table}[!h]
  \centering
  \caption{Representative per-unit computation times of the offline SF-profiling stages on a single NVIDIA RTX~3080 GPU (20\,GB).}
  \label{tab:sf-cost}
  \scriptsize
  \setlength{\tabcolsep}{4pt}
  \begin{tabular*}{\columnwidth}{@{\extracolsep{\fill}}l c@{}}
    \toprule
    Stage & Time per unit \\
    \midrule
    Scene segmentation (BaSSL) & 120\,ms / video second \\
    Frame captioning & 1\,s / frame \\
    Scene-level captioning & 3\,s / scene and version \\
    Scene weighting & 300\,ms / video \\
    Frame similarity & 100\,ms / frame pair \\
    Temporal judge & 160\,ms / scene pair \\
    \bottomrule
  \end{tabular*}
\end{table}

\subsection{Cross-MLLM stability of the Semantic Fidelity formulation}\label{app:cross-mllm}

To examine whether the MOS-correlation results reported in Section~\ref{sec:validation-empirical} depend on the choice of MLLM backbone, we repeat the SF evaluation using two additional open-source MLLMs of comparable size: Qwen2-VL-7B~\cite{wang2024qwen2vl} and InternVL2-8B~\cite{chen2024internvl2}. The two alternative MLLMs use different visual encoders and language backbones from VideoLLaMA3-7B. In all three configurations, we keep the prompt templates and the Qwen3-8B text-only model unchanged; only the MLLM used for frame captioning and scene summarization is replaced.

Table~\ref{tab:appB1} reports per-dataset SROCC and PLCC against MOS on LIVE-NFLX-II, SQoE-III, and SQoE-IV under the three backbones.

\begin{table}[!t]
  \centering
  \caption{Cross-MLLM SROCC and PLCC of Semantic Fidelity on three streaming-QoE datasets.}
  \label{tab:appB1}
  \footnotesize
  \setlength{\tabcolsep}{3pt}
  \begin{tabular*}{\columnwidth}{@{\extracolsep{\fill}}l cc cc cc@{}}
    \toprule
    \multirow{2}{*}{\textbf{MLLM Backbone}} & \multicolumn{2}{c}{\textbf{LIVE-NFLX-II}} & \multicolumn{2}{c}{\textbf{SQoE-III}} & \multicolumn{2}{c}{\textbf{SQoE-IV}} \\
    \cmidrule(lr){2-3} \cmidrule(lr){4-5} \cmidrule(lr){6-7}
    & SROCC & PLCC & SROCC & PLCC & SROCC & PLCC \\
    \midrule
    VideoLLaMA3-7B$^\dagger$ & \textbf{0.810} & 0.757 & \textbf{0.810} & 0.780 & \textbf{0.839} & 0.871 \\
    Qwen2-VL-7B          & 0.775 & 0.780 & 0.706 & 0.682 & 0.814 & 0.774 \\
    InternVL2-8B         & 0.846 & 0.844 & 0.723 & 0.706 & 0.804 & 0.777 \\
    \bottomrule
  \end{tabular*}
  \\[2pt]
  {\scriptsize $^\dagger$Main backbone used in Section~\ref{sec:validation}.}
\end{table}

Across the three datasets, SF shows similar correlations with MOS under all three MLLM backbones. This consistency indicates that the MOS-correlation results are not highly sensitive to the MLLM used for frame captioning and scene summarization.

\section{Closed-Form Analysis of the Buffer Controller}
\label{app:buffer-controller}

This appendix derives the buffer trajectory and settling-time estimate for the controller in~\eqref{eq:smoothing-controller} when $B_i+1-B_{\mathrm{th}}\geq 0$ and the playback buffer remains nonempty. When $B_i+1-B_{\mathrm{th}}\geq 0$, the max term equals $B_i+1-B_{\mathrm{th}}$, and the controller follows the linear recurrence analyzed below. We first assume that the actual delivery time equals its target and then analyze how the buffer changes when they differ for one batch.

\subsection{Buffer Trajectory without Delivery-Time Mismatch}

When $B_i+1-B_{\mathrm{th}}\geq 0$, the controller in~\eqref{eq:smoothing-controller} becomes
\begin{equation}
  \hat{D}_i=\lambda D_{i-1}
  +(1-\lambda)(B_i+1-B_{\mathrm{th}}),\quad i\geq1.
  \label{eq:app-controller}
\end{equation}
From the buffer recurrence~\eqref{eq:buffer-recurrence},
\begin{equation}
  D_i=-B_{i+1}+B_i+1,\qquad
  D_{i-1}=-B_i+B_{i-1}+1.
  \label{eq:app-buffer-duration}
\end{equation}
When $D_i=\hat{D}_i$, substituting~\eqref{eq:app-buffer-duration} into~\eqref{eq:app-controller} gives
\begin{align}
  -B_{i+1}+B_i+1
  &=\lambda(-B_i+B_{i-1}+1) \notag\\
  &\quad +(1-\lambda)(B_i+1-B_{\mathrm{th}}).
  \label{eq:app-controller-substitution}
\end{align}
Rearranging yields
\begin{equation}
  B_{i+1}-2\lambda B_i+\lambda B_{i-1}
  =(1-\lambda)B_{\mathrm{th}}.
  \label{eq:buffer-second-order}
\end{equation}

Let $\tilde{B}_i=B_i-B_{\mathrm{th}}$ be the buffer deviation. The recurrence becomes
\begin{equation}
  \tilde{B}_{i+1}-2\lambda\tilde{B}_i
  +\lambda\tilde{B}_{i-1}=0.
  \label{eq:app-nominal-deviation}
\end{equation}
Its characteristic equation is $z^2-2\lambda z+\lambda=0$, with roots
\begin{equation}
  z_{1,2}=\lambda\pm \mathrm{j}\sqrt{\lambda(1-\lambda)}
  =\sqrt{\lambda}\,e^{\pm \mathrm{j}\theta},
  \label{eq:app-controller-roots}
\end{equation}
where $\mathrm{j}^2=-1$, $\cos\theta=\sqrt{\lambda}$, and $\theta=\sin^{-1}\!\sqrt{1-\lambda}$. Hence,
\begin{equation}
  \tilde{B}_i=\lambda^{i/2}
  \left[c_1\cos(i\theta)+c_2\sin(i\theta)\right].
  \label{eq:app-general-buffer}
\end{equation}

With the initialization $\hat{D}_0=1$ and $D_0=\hat{D}_0$, we have $B_1=B_0$ and $\tilde{B}_1=\tilde{B}_0$. Applying these initial conditions to~\eqref{eq:app-general-buffer} gives $c_1=B_0-B_{\mathrm{th}}$ and $c_2=\sqrt{(1-\lambda)/\lambda}\,(B_0-B_{\mathrm{th}})$. For $i=0,1,2,\ldots$, the buffer trajectory is therefore
\begin{equation}
  \begin{aligned}
    B_i
    &=B_{\mathrm{th}}+(B_0-B_{\mathrm{th}})\lambda^{i/2}\\
    &\quad\times\left[
      \cos(i\theta)+\sqrt{\frac{1-\lambda}{\lambda}}\sin(i\theta)
    \right].
  \end{aligned}
  \label{eq:buffer-trajectory}
\end{equation}
Because both characteristic roots have magnitude $\sqrt{\lambda}<1$, $B_i$ converges to $B_{\mathrm{th}}$ for every $\lambda\in(0,1)$. The buffer deviation is bounded by $|B_0-B_{\mathrm{th}}|\lambda^{(i-1)/2}$. For $B_0\neq B_{\mathrm{th}}$, setting this bound to 5\% of the initial buffer deviation gives an estimate of the number of batches required for the buffer deviation to fall below this level:
\begin{equation}
  i_{0.05}\approx1+\frac{\ln 0.05}{0.5\ln\lambda}.
\end{equation}

\subsection{Effect of Delivery-Time Mismatch}

When $B_i+1-B_{\mathrm{th}}\geq0$, the actual delivery time may differ from its target because RB capacities are estimated and transmission occurs in discrete slots. Define the delivery-time mismatch as $w_i\triangleq\hat{D}_i-D_i$, so that $D_i=\hat{D}_i-w_i$. The buffer-deviation recurrence becomes
\begin{equation}
  \tilde{B}_{i+1}-2\lambda\tilde{B}_i
  +\lambda\tilde{B}_{i-1}=w_i.
  \label{eq:disturbance}
\end{equation}
Consider a unit impulse $w_i=\delta_i$, where $\delta_0=1$ and $\delta_i=0$ for $i\neq0$, under the zero initial conditions $\tilde{B}_0=\tilde{B}_{-1}=0$. Let $h_i$ be the resulting buffer-deviation response. Solving~\eqref{eq:disturbance} gives
\begin{equation}
  h_i=\frac{\lambda^{(i-1)/2}}{\sqrt{1-\lambda}}
  \sin\!\left(i\theta\right)u_{i-1},
  \quad \theta=\sin^{-1}\!\sqrt{1-\lambda},
  \label{eq:impulse-response}
\end{equation}
where $u_i=1$ for $i\geq0$ and $u_i=0$ otherwise. Since $D_i=1$ when the buffer remains at $B_{\mathrm{th}}$, the delivery-time deviation is $\tilde{D}_i\triangleq D_i-1=-h_{i+1}+h_i$. Using
\begin{equation}
  \sin((i+1)\theta)
  =\sqrt{\lambda}\sin(i\theta)
  +\sqrt{1-\lambda}\cos(i\theta)
\end{equation}
gives
\begin{equation}
  \tilde{D}_i=\lambda^{i/2}
  \left[
    \sqrt{\frac{1-\lambda}{\lambda}}\sin(i\theta)
    -\cos(i\theta)
  \right],
  \quad i\ge 0,\ \lambda\in(0,1).
  \label{eq:app-duration-response}
\end{equation}
Because the recurrence is linear, replacing $w_i=\delta_i$ with $w_i=a\delta_i$ multiplies both the buffer and delivery-time deviations by $a$. The factor $\sqrt{\lambda}$ determines how quickly these deviations decay, so the decay becomes slower as $\lambda$ approaches one.

\bibliographystyle{IEEEtran}
\bibliography{references}

\end{document}